\documentclass[preprint,review,10pt]{elsarticle}
\usepackage{bbm}
\usepackage{amsmath,amssymb,amsthm}
\usepackage[title]{appendix}
\usepackage{hyperref}
\usepackage{color}
\usepackage{xcolor}
\usepackage{graphicx}
\usepackage{enumerate}
\usepackage{ifpdf}
\usepackage{algorithmic}
\usepackage{subcaption}
\usepackage{float}
\usepackage{CJK}
\usepackage{listings}
\usepackage{bbm}
\usepackage{bm}
\usepackage{mathrsfs}
\usepackage{amsmath}
\usepackage{enumerate}
\usepackage{multirow,booktabs}
\usepackage{algorithm}
\usepackage{caption}
\usepackage{makecell}
\usepackage{epstopdf}
\usepackage{cases}
\usepackage{enumerate}
\usepackage{lipsum}
\usepackage{comment} 
\usepackage{suffix} % 或可能与计数器相关的其他宏包
\makeatletter
\def\ps@pprintTitle{%
	\let\@oddhead\@empty
	\let\@evenhead\@empty
	\def\@oddfoot{}%
	\let\@evenfoot\@oddfoot}
\makeatother
\usepackage{graphicx}
\usepackage{amssymb}
\usepackage{amsthm}

\makeatletter\@addtoreset{equation}{section} \makeatother
\usepackage[percent]{overpic}
\usepackage{tikz}
\usepackage{pgfplots} % <- Preamble
\usetikzlibrary{decorations.pathreplacing}
\usepackage{subcaption}
\usepackage{suffix} % 或可能与计数器相关的其他宏包
\include{Data_AdapEner}
\graphicspath{{Figures/}}

\usepackage{geometry}
\begin{document}
	\begin{frontmatter}
		\title{Adaptive Movement Sampling Physics-Informed Residual Network (AM-PIRN) for Solving Nonlinear Option Pricing models}
		%\tnotetext[label1]{This work is supported by National Natural Science Foundation of China (Grant No.12101310), Natural Science Foundation of Jiangsu Province (Grant No. BK20210315), 2021 Jiangsu Shuangchuang Talent Program (JSSCBS 20210222), the Fundamental Research Funds for the Central Universities (No. 30923010912).}
		
		\author{Qinjiao Gao\fnref{label1,label2}}
		\ead{qgao@mail.zjgsu.edu.cn}
            \author{Zuowei Wang\fnref{label1}}
		\ead{1615453100@qq.com} 
            \author{Ran Zhang\fnref{label3}*}
		\ead{zhang.ran@mail.shufe.edu.cn}	
            \author{Dongjiang Wang\fnref{label1}}
		\ead{18069175686@163.com}

		\address[label1]{School of Statistics and Mathematics, Zhejiang Gongshang University, Hangzhou, China}
		\address[label2]{ Collaborative Innovation Center of Statistical Data Engineering, Technology $\&$  Application, Zhejiang Gongshang University, Hangzhou, China}
		\address[label3]{School of Mathematics, Shanghai University of Finance and Economics, Shanghai, China}
		\cortext[cor1]{Corresponding author}
		\begin{abstract}

       In this paper, we propose the Adaptive Movement Sampling Physics-Informed Residual Network (AM-PIRN) to address challenges in solving nonlinear option pricing PDE models, where solutions often exhibit significant curvature or shock waves over time. The AM-PIRN architecture is designed to concurrently minimize PDE residuals and achieve high-fidelity option price approximations by dynamically redistributing training points based on evolving PDE residuals, while maintaining a fixed total number of points. To enhance stability and training efficiency, we integrate a ResNet backbone, replacing conventional fully connected neural networks used in Physics-Informed Neural Networks (PINNs). Numerical experiments across nonlinear option pricing models demonstrate that AM-PIRN outperforms PINN, RAM-PINN, and WAM-PINN in both resolving PDE constraints and accurately estimating option prices. The method’s advantages are particularly pronounced in complex or multi-dimensional models, where its adaptive sampling and robust architecture effectively mitigate challenges posed by sharp gradients and high nonlinearity.
            
			%In this paper, we introduce an adaptive sampling method named the adaptive ResNet-based Physically Informed Neural Network (PIRN) for solving Black-Scholes (BS) partial differential equations (PDEs). To better capture the physical properties of the BS equation, we make two significant enhancements to the conventional PINN framework. First, we implement a Residual-based Adaptive Refinement (RAR) strategy to improve solution accuracy, especially in regions with large variations or shock waves. Second, we enhance model stability by replacing the standard fully connected neural network in the PINN framework with a ResNet architecture. The proposed RAR-PIRN method exhibits robust performance in solving both nonlinear and standard BS equations. Numerical experiments confirm the effectiveness of this approach, particularly in tackling complex nonlinear BS equations.

		\end{abstract}
		
	\end{frontmatter}

	\label{Maintext}
	\section{\textbf{Introduction}}
   Partial differential equations (PDEs) are used to model various physical phenomena across different scientific fields, including physics, chemistry  and finance, etc. In financial markets, portfolio construction is a fundamental strategy for risk mitigation and plays a crucial role in determining the pricing dynamics of financial assets. Typically, deterministic methods are employed for option pricing by solving the Kolmogorov PDE based on the Feynman-Kac formula. A cornerstone of option pricing theory is the Black-Scholes (BS) equation \cite{black1973pricing}, which provides a fundamental framework for pricing financial derivative contracts. 
     
     However, in Black-Scholes equations, the underlying assets are following the geometric Brownian motion (GBM),  which often fail to capture the complexities and nuances of real-world market behavior. As a result, researchers have turned their attention to more sophisticated models that incorporate features like stochastic volatility, jumps in asset prices, and changing correlations among assets. For instance, Merton jump-diffusion model \cite{MERTON1976125}, the Constant Elasticity of Variance (CEV) \cite{Cox}, the Heston stochastic volatility models \cite{Heston}, and rough Bergomi model under rough volatility \cite{Bayer02062016}, etc.
   
    Analytical solutions to option pricing equations are often intractable for realistic financial models, prompting the need for advanced numerical methods. Unlike stochastic approaches like Monte Carlo simulations that rely on random sampling, traditional deterministic numerical methods provide grid-based precision and guaranteed convergence for low-to-moderate dimensional problems. For example, Koffi and Tambue \cite{koffi2020fitted} introduced a specialized finite volume method tailored for the BS model  that ensures stability and convergence in high-dimensional settings.  Hon  \cite{hon1999radial} developed the radial basis function finite difference (RBF-FD) method to solve the BS equation,  while Zhang \cite{zhang2017radial} proposed a multinomial tree approach. Dillon and Tangman \cite{dilloo2017high} developed a higher-order finite difference method for various option pricing models.
 
    The integration of deep learning into financial mathematics has opened new frontiers for solving option pricing PDEs, particularly in high-dimensional or data-driven settings.  Raissi et al. \cite{raissi2019physics} introduced physics-informed neural networks (PINNs), that embeds PDE constraints directly into neural network training to solve both forward pricing problems and inverse parameter calibration tasks. Wang et al. \cite{wang2023deep} demonstrated the effectiveness of PINNs for both linear and nonlinear Black-Scholes (BS) equations, highlighting their simplicity, accuracy, and computational efficiency. In \cite{fBS},  PINN is applied to solve time fractional Black-Scholes equations.  The Laguerre neural network was proposed as a novel numerical algorithm with three layers of neurons for solving the generalized BS equations \cite{chen2021numerical}. 
    %Bae et al. \cite{bae2024option} further explored the application of PINNs in option pricing and local volatility modeling. 
    
   Despite some empirical successes, PINNs continue to face several challenges, highlighting open areas for research and further methodological advancements. Numerous studies have aimed at enhancing the performance of PINNs, primarily by developing more effective neural network architectures and improved training algorithms  \cite{WANG2022110768, YU2022114823, CHIU2022114909,DAS,gao2023failure}.
   PINNs are primarily optimized based on the PDE loss, which ensures that the trained network aligns with the PDE being solved. This loss is evaluated at a set of scattered residual collocation points, similar to the grid points used in traditional numerical methods. Therefore, the location and distribution of these residual points are crucial for the performance of PINNs. It has been shown theoretically and experimentally that PINNs often struggle to learn high-frequency features \cite{frequencybias}. 
   
   As a result, instead of using the same residual points throughout the training process, we could select a new set of residual points during optimization iterations. This idea is particularly important given that solutions to financial pricing equations can exhibit significant curvature or even shock waves over time, even when originating from smooth initial conditions \cite{jump}. Besides, option pricing equations are often highly nonlinear, especially when considering path-dependent options (like Asian options or barrier options). This nonlinearity can lead to significant fluctuations in the solution in specific regions.

    The residual-based adaptive refinement method (RAR), as described in \cite{lu2021deepxde}, is the most common strategy. RAR operates by continuously adding points with larger residuals to the collocation point set, which enhances training efficiency and accuracy.  In Residual-based adaptive refinement with distribution (RAD) \cite{wu2023comprehensive}, all residual points are resampled based on a probability density function proportional to the PDE residual, aiming for a more balanced distribution of attention across the domain. Building on this, residual-based adaptive refinement with distribution (RAR-D) combines the approaches of RAR-G and RAD. While these strategies effectively reduce the PDE residuals by concentrating on regions with larger errors, they may inadvertently introduce greater function errors and lead to an uneven distribution of function errors across the domain. Furthermore, the process of adding more collocation points during training inevitably increases computational time. 
    \cite{AW} relies on monitor functions \cite{GAO2018115} to guide collocation point adjustments. However, in practice, these monitor functions, which are widely used in classical adaptive methods, often necessitate a handcrafted design. 
    %Residual-based adaptive distribution (RAD) and residual-based adaptive refinement with distribution (RAR-D) \cite{gao2023failure,noorizadegan2024power,chen2023adaptive} improve computational efficiency by focusing resources on regions of high curvature, refining the solution process. %While RAR shows promise in improving accuracy through adaptive sampling, our findings reveal limitations in certain complex scenarios, suggesting the need for further enhancements.

   In this paper, we introduce the Adaptive movement sampling Physics-informed residual networks (AM-PIRN), whose architecture is designed to simultaneously minimize the PDE residuals and achieve high-fidelity function approximation. 
   It integrates $r$-adaptation with deep learning for solving nonlinear option pricing equations.  The approach preserves the total number of points while dynamically redistributing the movable points during training based on the PDE residuals. This movement is triggered at selected training steps. 
   %our method directly leverages the PDE residual as a driving mechanism for adaptive sampling, eliminating the need for handcrafted monitor functions.
    To enhance stability and training efficiency in adaptive refinement scenarios, we replace the fully connected neural networks employed in PINNs with a ResNet backbone \cite{he2016deep}. ResNet’s skip connections mitigate gradient degradation and enable robust function feature learning. 
    
    %This integration ensures accurate solution estimation while maintaining convergence stability, even under dynamic changes in the collocation point distribution. 

   %However, minimizing the target loss function does not necessarily lead to a corresponding reduction in function error. Then we still cannot obtain an accurate option price estimation. 
  %we propose a novel framework, AM-PIRN, which integrates ResNet architectures \cite{he2016deep} into the traditional PINN framework, replacing the conventional fully connected neural networks. The RAR-PIRN method is designed to improve numerical stability and computational efficiency, particularly when dealing with steep gradients in time-dependent PDEs. Our approach shows strong performance in solving both nonlinear and general BS equations, with numerical experiments validating the effectiveness of the RAR-PIRN method for tackling complex BS equations.

   The structure of this paper is as follows: Section 2 introduces the foundational mathematical framework for option pricing and details the architecture of PINNs. Section 3 presents the core methodological innovations of this work. We propose the integration of an adaptive collocation point movement strategy into the Physics-Informed Resnet Network (PIRN) framework, followed by a systematic exposition of the algorithm’s design. Section 4 demonstrates numerical validation through experiments across diverse option pricing models to assess the efficacy and robustness of our approach. Finally, Section 5 summarizes our findings and outlines directions for future research.
  	
\section {\textit{Preliminary}}
\subsection{The Option Pricing PDEs}

Firstly, we will introduce the mathematical models for financial derivatives such as options or futures. In order to price any derivative, it is essential to model the statistical behavior of the underlying asset at every point leading up to the expiry date. This involves describing the evolution of an asset's price using a stochastic process. Various stochastic models for asset prices exist to accurately price different derivatives. 
%These results are typically based on stochastic processes and probability theory.  
%we will introduce the Feynman-Kac formula in probability theory and mathematical finance, which connects stochastic processes, particularly Brownian motion, with partial differential equations (PDEs). 

Considering that the underlying asset is a stochastic process $S_t \in \mathbb{R}^d \  (t \in [0,T])$ described by the following It$\hat{o}$-type SDEs
\begin{equation}
    dS_t=\mu(S_t,t)dt + \sigma(S_t,t)dW_t, %S(0)=S_0,
\end{equation}
where the Borel-measurable functions $\mu(S_t,t): \mathbb{R}^d\times [0,T] \rightarrow \mathbb{R}^d$ and $\sigma(S_t,t): \mathbb{R}^d\times [0,T] \rightarrow \mathbb{R}^{d \times d}$  being the drift term and  the volatility term respectively, and $W_t$ is a standard Brownian motion. The SDEs are well-defined only with certain boundary conditions. 
Here $T \in (0,+\infty)$ is the maturity.

Let $\psi(S): \mathbb{R}^d \rightarrow \mathbb{R}$ be the payoff function and $U=U(S,t) \in C^{2,1}\left(\mathbb{R}^d \times [0, T]\right)$ being the option price function or financial derivatives such that $U(S,T)=\psi(S)$,  and it holds for all $t \in[0, T)$ and $ S \in \mathbb{R}^d$

\begin{equation*}
U(S,t)=\mathbb{E}\left[U\left(S_{T},T\right) \mid S_{t}=S\right]=\mathbb{E}\left[\psi\left(S_{T}\right) \mid S_{t}=S\right].
\end{equation*}
%With $\tau=T-t$, we define the function $U(S,\tau)=V(S,T-t)$ satisfying $U(S,0)=\psi(S)$. In the following text, without causing ambiguity, we  denote  $\tau$ as $t$.
Then from It$\hat{o}$'s rule \cite{ito}, we have
\begin{equation}\label{Ito}
dU_t= \left(\frac{\partial U}{\partial t}+\frac{1}{2}\text{Tr}(\sigma \sigma^T \nabla_S^2 U)+\mu \cdotp \nabla_S U\right)dt+ \nabla_S U \cdotp \sigma \cdotp dW_t.
\end{equation}
According to the risk-neural pricing or arbitrage pricing theory \cite{risk}, Equ.(\ref{Ito}) is modified to accommodate the risk-free interest rate function $r(S_t,t): \mathbb{R}^d\times [0,T] \rightarrow \mathbb{R}$, then we have the corresponding partial differential equation
\begin{equation}\label{BS}
    \frac{\partial U}{\partial t}+\frac{1}{2}\text{Tr}(\sigma \sigma^T \nabla_S^2 U)+ rS \cdotp \nabla_S U -rU=0
\end{equation}
with  finial payoff condition $U(S,T)=\psi(S)$. 
Here $\nabla_S U$ and $\nabla_S^2 U$ denote the gradient and the Hessian of
function $U$ with respect to $S$, and $Tr$ denotes the trace of a matrix. We are most interested in the solution at $t = 0$, i.e.,  the option  premium $U(S,0)$.
With $t \leftarrow T-t$, we can easily define the function $U(S,t)$ satisfying the initial condition $U(S,0)=\psi(S)$.  One can also use the Feynman-Kac formula \cite{ito} to establish a connection between partial differential equations and stochastic processes, then to derive the option pricing equation. Typical choices of the payoff function includes the European options, American options, and other path-dependent options such as Asian or barrier options. In this paper, to simplify the modeling process, we only consider the payoff which is determined solely at expiration.

In the vanilla Black-Sholes model, $S_t$ follows the geometric Brownian motion and Equ.(\ref{BS}) is reduced to a linear PDE. 
In the following content,  the results of the general nonlinear cases will be studied. 
We will focus on the Barles' and Soner's model in \cite{hodges1989optimal}, where the transaction cost is considered. The volatility $\sigma$ depends on the time $t$, the asset $S$ or the derivatives of the option price $U$ itself. 
The model can be derived by an exponential utility function  and subsequently refined by Davis et al. \cite{davis1993european}. The existence and uniqueness of the solution of the problem were shown in the context of stochastic optimal theory in \cite{barles1998option}. 
In addition, the constant elasticity of variance model (CEV) introduced  by Cox\cite{Cox}, which uses $\sigma S_t^{\beta/2}$ ($\sigma$ being the constant volatility) as the local volatility. The closed-form solution of CEV has been provided for European call option.  We will consider the Heston Stochastic Volatility Model \cite{Heston} which extends the Black-Scholes equation by allowing volatility to follow a CIR-process. Its ability to capture the volatility smile and skew in implied volatility makes it a valuable tool for practitioners in finance.  Heston in \cite{Heston} derived a closed-form solution for the price of a European call option using the model's characteristic function, however the integral in the solution requires numerical computation.

\begin{comment}
    The theory of option pricing has progressed considerably since the inception of the Black-Scholes option pricing model. The renowned Black-Scholes equation is expressed as follows: 
	\begin{align}\label{eq:1}
		\begin{split}
			U_t+\frac{1}{2}\sigma ^2S^2U_{SS}+rSU_S-rU=0, on\,\, \left( 0,\infty \right) \times \left( 0,T \right] .
		\end{split}
	\end{align}
	where the parameters $r$ and $\sigma$ signify the risk-free rate and volatility, respectively. The equation denotes the fair value of a vanilla European option with an asset value $S$ at time $\tau$. Here, $T$ designates the expiration date of the contract, $\tau=T-t$ represents the time to expiration, and $t \in \left( 0, T \right]$ signifies the instantaneous time.
	The volatility $\sigma$ may be contingent on the derivative of time ($t$), the stock price ($S$), or the option price ($U$) itself. In such cases, (\ref{eq:1}) transforms into the following nonlinear Black-Scholes equation.
	\begin{align}\label{eq:2}
		\begin{split}
			U_t+\frac{1}{2}\widetilde{\sigma }^2\left( t,S,U_S,U_{SS} \right) S^2U_{SS}+rSU_S-rU=0, on\,\, \left( 0,\infty \right) \times \left( 0,T \right]  .
		\end{split}
	\end{align}
\end{comment}

Given the complexity of many derivatives, closed-form solutions are not always practical. Instead, numerical methods are frequently employed to estimate prices by simulating the underlying asset's price paths. The non-deterministric methods such as Monte-Carlo simulations and the related modified sampling methods usually converge slowly and time-consuming. Meanwhile, the deterministic methods \cite{com_fin} like finite difference methods, and binomial trees suffer from the curse of dimensionality. 
Importantly, most numerical methods are tailored for specific models, making them less flexible when adapting to new or complex derivatives. They can be highly sensitive to input parameters, which may lead to instability in results if parameters are estimated poorly. 

%{\color{red} inverse problem}

	\subsection{Physics-informed neural network (PINN)}
    Based on the above discussion, 
	now we will provide a succinct introduction to PINN to solve the nonlinear PDEs  (\ref{BS}). Especially, for high-dimensional data, it is capable of capturing more complex nonlinear relationships and providing fast computation in many fields. For more details, one can refer \cite{high_dim, DGM}.      
    
    Consideration is given to a generic nonlinear partial differential equation, 
 	\begin{align}\label{eq:8}
		\begin{split}
			& \frac{\partial U}{\partial t}+\mathcal{N}\left[ U;\lambda \right]=0,    (S,t) \in \varOmega \times \left( 0,T \right] ,\\
			&u\left(S,0 \right) = u_0(S) ,        S \in  \varOmega ,\\
			&u\left( S,t \right) =B\left(S,t \right) ,  (S,t) \in  \partial \varOmega \times  \left[ 0,T \right],
		\end{split}
	\end{align}
	where $U(S,t)$ is the hidden solution as mentioned above, $\mathcal{N}$ is a nonlinear differential operator acting on $U(S,t)$ with respect to the variable $S$ with parameters $\lambda$, and $\varOmega$ is a subset of $\mathbb{R}^d$. 
We assume operators $\mathcal{N}$, initial and boundary conditions are known, and $\lambda$ 
in the governing equation is a known constant in the forward problem. For simplicity,  the final condition in the option pricing equations is also called the initial condition.  
In the following context, we will denote the underlying asset variable $S$ and $t$ as $x=(S,t)$.
% In this paper, we will focus on $d=1$. 
    
    %In the formula provided, $x$ and $t$ signify the spatial coordinates and time coordinates, respectively. $u_t$ represents the time derivative term, and $\mathcal{N}_\textbf{x}\left[ \cdot \right]$ is a differential operator. In PINNs, the treatment of the initial condition aligns with how Dirichlet boundary conditions are handled.
	
	In \cite{raissi2019physics}, Raissi etc. employ a fully connected neural network $f\left( x;\Theta \right)$ to approximate the solution of the function $U(x)$, where $\Theta$ represents the parameter set of the neural network. The deep neural network (DNN) $f\left(x;\Theta \right)$ with $L-1$ hidden layers, and an output layer is denoted as follows:
	
	\begin{align}\label{eq:9}
		\begin{split}
			&x^0=x^T,               \\
			&\mathscr{L}_l =\phi( w^{\left( l \right)}x^{l-1}+b^{\left( l \right)}),\ l=1,\cdots,L-1,\\
            &f\left( x;\Theta \right):=\mathscr{L}_L= w^{\left( L \right)}x^{L-1}+b^{\left( L \right)},
			%&f\left(x,t;\theta \right) =\left( \mathscr{L} _L\cdots 
			%\sigma \mathscr{L} _1 \right) \left( x^0 \right),  
		\end{split}
	\end{align}
where $\mathscr{L}_l$ denotes the output of the $l$-th layer, and $\mathscr{L}_L$ is the output of the neural network. $\phi(\cdot)$ denotes the activation function, which allows a neural network to map nonlinear relationship.  $\Theta=\{w^{\left( l \right)}, b^{\left( l \right)}\}_{l=1}^L$ represents the weight and bias of all the $L$ layers, which will be updated during training. 
The automatic differentiation is utilized to calculate the derivatives of the DNN $f\left( x;\Theta \right)$ with respect to $S$ and $t$ by the backward chain rule.

Based on machine learning framework, PINN effectively integrate physical information with deep neural networks. %A typical PINN framework for solving the Equ.(\ref{eq:8}) is depicted in Fig.(\ref{fig1}). 
Denote the input training data as $\mathcal{X}$ which includes the sampling points $\mathcal{X}_i$ on the initial condition, the sampling points $\mathcal{X}_b$ on the boundary condition, and collocation points $\mathcal{X}_p$ in the equation domain. In training process, they are usually randomly distributed points in the domain. The approximation function $f\left(x;\Theta \right)$ i.e., the  parameter $\Theta$ is learned with the input points by minimizing the loss function

	\begin{align}\label{eq:11}
		\begin{split}
			&\mathcal{L} \left( \Theta ;\mathcal{X}\right) =
            \omega _p\mathcal{L} _p\left( \Theta ;\mathcal{X}_p \right) +
            \omega _b\mathcal{L} _b\left( \Theta ;\mathcal{X}_b \right) +\omega _i\mathcal{L} _i\left( \Theta ;\mathcal{X}_i\right),
		\end{split}
	\end{align}
	where
	\begin{align}\label{eq:12}
		\begin{split}
			&\mathcal{L} _p\left( \Theta ;\mathcal{X} _p \right) =\frac{1}{\left| \mathcal{X} _p \right|}\sum_{x\in \mathcal{X} _p}{\left| r\left(x;\Theta \right) \right|^2},\\ 
			&\mathcal{L}_b\left( \Theta ;\mathcal{X} _b \right) =\frac{1}{\left| \mathcal{X}_b \right|}\sum_{x\in \mathcal{X} _b}{\left| f\left( x;\Theta \right) -B(x) \right|^2},\\
			&\mathcal{L} _i\left( \Theta ;\mathcal{X}_i \right) =\frac{1}{\left| \mathcal{X} _i \right|}\sum_{S\in \mathcal{X} _i}{\left| f\left( S,0;\Theta \right) -u_0\left( S \right) \right|^2},	
		\end{split}
	\end{align}
in which the PDE residual is defined as
	\begin{align}\label{eq:10}
		\begin{split}
			&r\left(x;\Theta \right) :=\frac{\partial}{\partial t}f\left(x;\Theta \right) +\mathcal{N}\left[ f\left(x;\Theta \right) \right].
		\end{split}
	\end{align}
Here $\omega_f$, $\omega_b$, and $\omega_i$ denote the positive scalar weights. Thus PINNs leverage the inherent constraints and boundary conditions embedded in partial differential equations to expedite network training. 
    %Specifically, PINNs integrate three essential sets of data points: one within the domain ($\mathcal{T}_f$), two on the boundary ($\mathcal{T}_b$), and the initial conditions ($\mathcal{T}_i$). The formulation of the loss function is delineated as follows:
Besides, it is shown that the boundary conditions can be precisely and adaptively enforced by adjusting the network structure \cite{krishnapriyan2021characterizing, bischof2021multi}, thereby we can eliminate the loss term of the boundary conditions.

%{\color{red}
%For solution to the inverse problem of PDEs, i.e., to discover the unknown parameters $\lambda$ in the PDE with measurement data, only minor modifications to the PINN framework.
%}

%After preparing the training data, the network's parameters are optimized using gradient descent methods commonly applied in deep learning, such as Adam, SGD, or L-BFGS. By minimizing the loss function, the neural network output approximates the solution function $U(S,t)$ of (\ref{eq:8}) when the loss reaches its minimum.

\section{\textit{Adaptive movement sampling Physics-informed residual network (AM-PIRN)}}

\subsection{Methodology}

Physics-informed neural networks (PINNs) utilize fully connected neural networks, but they encounter significant challenges. Particularly in the problems with complex differential equations. 
In this section, we propose Adaptive movement sampling Physics-informed residual network (AM-PIRN) shown in Fig. (\ref{fig5}). 
%We utilize the adaptive collocation point method in \cite{AW} to improve the distribution of residual points during training process. 
Unlike RAR and RAD, where all collocation points are updated during the adaptive process,  inspired by \cite{AW}, we separates the collocation points in two sets, i.e., the fixed set and the movable set. It focuses solely on updating the movable collocation points  during training.  
However, minimizing the target loss function does not necessarily lead to a corresponding reduction in function error. Then we still cannot obtain an accurate option price estimation. 
\cite{AW} suggests to rely on monitor functions to guide collocation point
adjustments. However, in practice, these monitor functions often need a handcrafted design. So we keep to resample the movable points following the distribution associated with the PDE residuals, and at meantime, the solution characteristics are taken into account in the network architecture. We replace the fully connected neural networks employed in PINNs with a ResNet backbone \cite{he2016deep}. ResNet’s skip connections mitigate gradient degradation and enable robust function feature learning. 

%Drawing inspiration from the improved numerical stability and computational efficiency of ResNet over conventional methods, and recognizing that ResNet's residual structure aids gradient propagation—especially in adaptive refinement scenarios.

%Given the absence of a more effective error representation metric than residuals, our study focuses on improving model performance by modifying the network structure. In our proposed {\bf{Algorithm 1}}, the traditional fully connected neural network in PINNs is replaced to enhance model stability. Drawing inspiration from the improved numerical stability and computational efficiency of ResNet over conventional methods, and recognizing that ResNet's residual structure aids gradient propagation—especially in adaptive refinement scenarios—we introduce a new approach, termed RAR-PIRN. The structural diagram is provided in Fig. \ref{fig5}.

%Residual-based adaptive sampling in PINNs, though intended to address such challenges, exhibits similar limitations in the case of the Klein-Gordon equation. While increasing training points in regions with large residuals can reduce the residual, this approach may inadvertently introduce greater errors and result in uneven error distribution. This highlights the fact that minimizing residuals does not necessarily lead to a corresponding reduction in error. 

%such as parameter redundancy and convergence issues. For more complex differential equations, these networks may struggle to effectively capture the underlying physical information. Furthermore, 

\subsection {\textit{Algorithm}}

We firstly provide the adaptive sampling strategy.   	
From a prior probability density function $p_0(x)$ (e.g., the uniform distribution), we initially sample the sets of collocation points in the equation domain as $\mathcal{X}_p$, 
the  points  on the initial condition as $\mathcal{X}_i$, the points on the boundary condition as $\mathcal{X}_b$. 
In the training process,  
the former set  then will be adjusted through adaptive sampling to optimize the model's performance and training efficiency, while the latter two sets remain unchanged.  
%$P_{N_{i}}$, the set of boundary points as $P_{N_{b}}$, and the set of $N_{r}$ collocation points as $P_{N_{r}}$. Therefore,

%The set $P_{N_{rn}}=\left{ x_{rn}^{j},t_{rn}^{j} \right} {j=1}^{N{rn}}$ represents a fixed collection of collocation points, uniformly sampled and constant throughout the training process. In contrast, $P_{N_{rm}}=\left{ x_{rm}^{j},t_{rm}^{j} \right} {j=1}^{N{rm}}$ consists of mobile collocation points that adaptively move toward significant computational regions as training progresses. This movement mimics the behavior of the moving grid method, where points converge towards areas with pronounced solution gradients. This dynamic adjustment is achieved by resampling $p(x)$ in the probability density function (\ref{eq:15}), guiding the collocation points towards critical computational regions.

%Follow \cite{gao2023failure}, 
%we firstly provide the limit-state function (LSF) $ g(x): X \to  \mathbb{R}$, i.e.,
%$$g(x)= Q(x) - \delta.$$ 
%Here, $\delta$ is a predefined maximum allowed threshold, and $Q(x): x \to R$ maps the domain
%to a quantity of interest (QoI) that characterizes the system’s performance. Here we
%can choose $Q(x) = |p(x; \Theta)|$ with $p(x; \Theta)$ being the probability density function. It is defined by 
Following the adaptive collocation point method in \cite{AW}, $\mathcal{X}_p$ is divided into two sets, i.e.,
$$\mathcal{X}_p = \mathcal{X}_{p,i} \cup \mathcal{X}_{p,m}$$
where $\mathcal{X}_{p,i}$ is the immovable set of collocation points which do not change throughout the training of a model. $\mathcal{X}_{p,m}$ refers a collection of movable points that gradually converge on the key computational areas (e.g., where shock waves are generated) during the training process, akin to the grid movement in the moving mesh method. 
Thus, by integrating $\mathcal{X}_{p,i}$ and $\mathcal{X}_{p,m}$, the collocation points can achieve adaptive movement, maintaining a uniform distribution in smooth areas while concentrating more densely in challenging regions. This approach allows for customization to specific problems and automatically adjusts the distribution of collocation points, enhancing model accuracy.

The movement of the movable collocation points  $\mathcal{X}_{p,m}$ 
is accomplished through resampling according to the probability density function $p(x; \Theta)$. 
It is defined as 

    \begin{align}\label{eq:15}
		\begin{split}
			p(x; \Theta) \propto \frac{|r^k\left(x;\Theta \right)|}{\mathbb{E}[{|r^k\left(x;\Theta \right)}|]}, %\ \ x \in \mathcal{X}_p,
		\end{split}
	\end{align}
with $k\geq 0$ being the power hyperparameter and $r(x,\Theta)$ given in Equ.(\ref{eq:10}).

Specifically, in selected training steps, we can generate a
set of randomly distributed locations $\mathcal{X}^0_{p,m} = \{x_1, x_2,\dots, x_{|\mathcal{X}^0_{p,m}|}\}$ from the prior $p_0(x)$.
Then we approximate $\mathbb{E}[{|r^k\left(x;\Theta \right)}|]$ in (\ref{eq:15}) by Monte Carlo integration, then it is
\begin{align}\label{p}
		\begin{split}
			\tilde{p}(x_i; \Theta) =\frac{|r^k\left(x_i;\Theta \right)|}{\sum_{x_i \in \mathcal{X}^0_{p,m}}{|r^k\left(x_i;\Theta \right)}|}, \ \ x_i \in \mathcal{X}^0_{p,m}.
		\end{split}
	\end{align}
In the end, a dense subset $\mathcal{X}_{p,m}$ of $M$ points from $\mathcal{X}^0_{p,m}$ can be resampled according to the discrete density distribution $\tilde{p}(x_i;\Theta)$. The new collocation points will be updated to $\mathcal{X}_p = \mathcal{X}_{p,i} \cup \mathcal{X}_{p,m}$, then we can retrain the network.
%compute $p(x_i;\Theta)$, 

%Then the hypersurface defined by $g(x) = 0$ divides the physics domain into two subsets:
%the safe set $ X_s= \{x : g(x) < 0\}$ and the update set $ X_u = \{x : g(x) > 0\}$. 
%Be specifically,  initially we may generate a
%set of randomly distributed locations $\mathcal{X}_p = \{x_1, x_2,\dots, x_{|\mathcal{X}_p|}\}$ from the prior $p_0(x)$ (e.g., the uniform distribution), 
%then we can conduct the adaptive sampling method by computing $p(x_i;\Theta)$. If 

It can be noticed that the hyperparameter $k$ plays a crucial role in this process. When $k=0$, $p(x;\Theta)$ is approximated by a uniform distribution, spreading the sampling points evenly across the computational domain. As $k$ increases, only points
with the largest PDE residual are selected. The combination of fixed and adaptive collocation points ensures  that the model focuses on key areas during training, enhancing efficiency, and ultimately improving overall performance.
In fact, this idea is the moving grid method, where points converge towards areas with   steep gradients of the solution of the equations. %This dynamic adjustment is achieved by resampling $p(x)$ in the probability density function (\ref{eq:15}), guiding the collocation points towards critical computational regions.

As aforementioned, resampling in PINNs primarily focuses on minimizing the equation loss in the domain, however, this approach does not directly address the solution error. So we introduce ResNet to specifically control the solution error in the training, which allows us to incorporate skip connections that help retain important features and improve the model's ability to learn complex functions.  

\begin{figure*}[t!]
		\centering
		\includegraphics[width=0.8\textwidth]{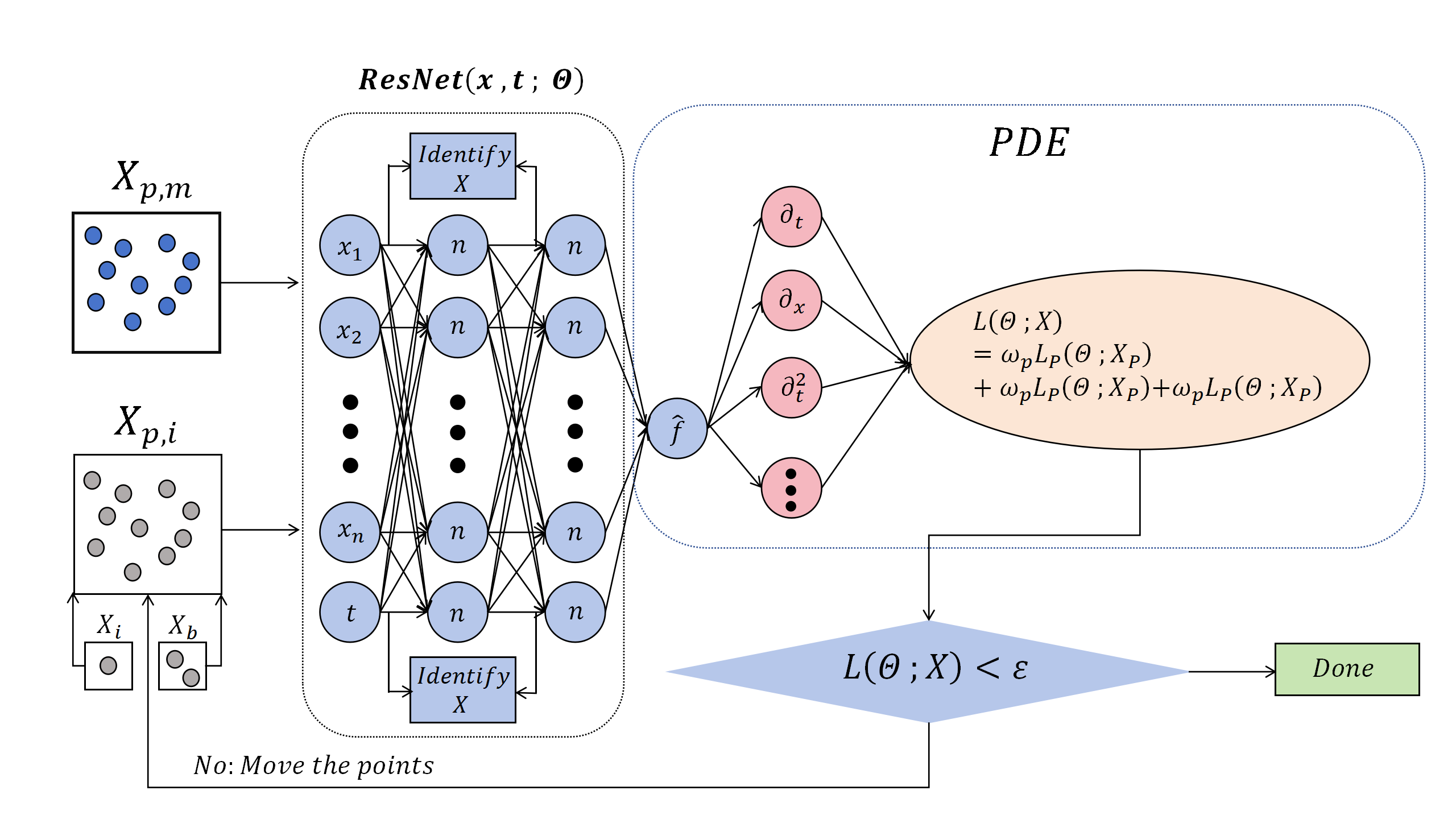}     
		\caption{The workflow of AM-PIRN: Adaptive movement sampling Physics-informed residual network. $ \mathcal{X}_{p,i}$ is the fixed set of collocation points that remains constant during the model training, generated using a uniform sampling strategy. $\mathcal{X}_{p,m}$ is the set of movable collocation points that will dynamically change during the training process.}
		\label{fig5}
	\end{figure*}

Suppose that we intend to learn the underlying mapping $H(x)$ by a few stacked layers,  which is not necessary to apply to the entire net. 
In our problem, $H(x)$ would be the PDE solution function $U(x)$. 
In ResNet, the problem is transformed into learning a residual function $F(x) := H(x)-x$, then the original output function becomes $F(x)+x$. A building block can be
demonstrated as

\begin{equation}\label{Resnet}
    U=F(x,\{w^{(i)}\})+w^sx, 
\end{equation}
where $x$, $U$ denotes input and output values, respectively and the function $F(x,\{w^{(i)}\})$ is the residual mapping to be trained. The matrix $w^s$ is only used to match dimensions. For example, 
a two layers network can be expressed as $F = w^{(2)}\phi(w^{(1)}x)$, where $\phi$ is the activation function and the biases are omitted for simplifying notations. 
Then by bypassing the activation function, the output from one or more preceding layers is added to the current layer's output before being fed into the activation function, which then serves as the output for the current layer. The whole architecture can be found in Fig. \ref{fig5}.

	\begin{algorithm}[h]
		\caption{Solving Option pricing Equations by AM-PIRN}
		\label{alg:AOS}
		\renewcommand{\algorithmicrequire}{\textbf{Input:}}
		\renewcommand{\algorithmicensure}{\textbf{Output:}}
		\begin{algorithmic}[1]
			\REQUIRE  Maximum iterations $N_{max}$,  $|\mathcal{X}_p|$,  $|\mathcal{X}_i|$, $|\mathcal{X}_b|$, the number of updated points $M$, the tolerance of the residuals $\epsilon$;
            %$N_{rn}$ , $N_{rm}$ , the nonlinear Black-Scholes equation , RAR-count.  %%input
			\ENSURE $\Theta^*$    %%output
            \STATE  Sample the initial   $\mathcal{X}_p$,  $\mathcal{X}_i$, and $\mathcal{X}_b$ with the prior uniform density function $p_0(x)$;
            \REPEAT 
			\STATE Train ResNet $f(x;\Theta)$ with the training dataset $\mathcal{X}_{p}$,  $\mathcal{X}_i$ and $\mathcal{X}_b$ for a certain number of iterations;
            \STATE Sample a set of dense points $\mathcal{X}^0_{p,m}$ using $p_0(x)$;
			\STATE Compute density function $\tilde{p}(x_i;\Theta)$, $x_i \in $ $\mathcal{X}^0_{p,m}$ with (\ref{p});
            \STATE Resample $M$ points from $\mathcal{X}^0_{p,m}$ with $\tilde{p}(x_i;\Theta)$ and get $\mathcal{X}_{p,m}$;
            \STATE Update $\mathcal{X}_p \leftarrow \mathcal{X}_{p,i} \cup \mathcal{X}_{p,m}$;
			%\IF {$\text{error} < \epsilon$}
			%\STATE \textbf{break}
			%\ENDIF %{$i=1,\cdots, N_{max}$}
            \UNTIL{$\mathcal{L}\left( \Theta ;\mathcal{X}\right) < \epsilon$ or the total number of iterations larger than $N_{max}$;}
			\RETURN $\Theta^*$.
		\end{algorithmic}
	\end{algorithm}
  
 %Given the absence of a more effective error representation metric than residuals, our study focuses on improving model performance by modifying the network structure. In our proposed {\bf{Algorithm 1}}, the traditional fully connected neural network in PINNs is replaced to enhance model stability. Drawing inspiration from the improved numerical stability and computational efficiency of ResNet over conventional methods, and recognizing that ResNet's residual structure aids gradient propagation—especially in adaptive refinement scenarios—we introduce a new approach, termed RAR-PIRN. The structural diagram is provided in Fig. \ref{fig5}.
 
 \section {\textit{Numerical simulations}}
	In this section, the effectiveness of the proposed AM-PIRN method is substantiated by the following numerical experiments. The code of all the examples are implemented by Pytorch. 
    
    We will compare AM-PIRN proposed in this paper with PINN \cite{raissi2019physics}, the adaptive collocation point method for PINN based on residuals  (RAM-PINN) 
    and monitor function (WAM-PINN) \cite{AW} in the following examples. Specifically, in WAM-PINN the resampling density function is defined by
       \begin{align}\label{eq:WAM}
		\begin{split}
			p(x; \Theta) \propto \frac{|m^k\left(x;\Theta \right)|}{\mathbb{E}[{|m^k\left(x;\Theta \right)}|]}, %\ \ x \in \mathcal{X}_p,
		\end{split}
	\end{align}
    with the monitor function defined by
    $$
    m(x)=\sqrt{1+\|\nabla_x U\|_2^2}.
    $$

    To ensure a fair comparison of different models, all comparative experiments were executed under the same conditions.  Adam-iters (number of Adam iterations),  L-BFGS-iters (maximum number of iterations for L-BFGS), Adam-lr/L-BFGS-lr (Adam  and LBFGS learning rate), and parameter $k$ in the probability density function $\tilde{p}(x;\Theta)$ are listed in Table (\ref{Table:para}).  The fully connected network used in PINN contains five hidden layers with 20 neurons in each hidden layer, while the residual function $F(x,\{w^{(i)}\})$ in Equ.(\ref{Resnet}) has two hidden layers with 20 neurons in each hidden layer. The Tanh activation function is employed in all the trainings. The weights of the loss function are set to be $\omega_p=\omega_b=\omega_i=1$. The collocation points are initialized as uniformly sampled points within the computational domain for all the methods. The number of iteration rounds for RAM-PINN, WAM-PINN and AM-PIRN are all set to be $10$. All the hyper-parameters were maintained consistently in all experiments, unless stated otherwise.
    %Two error functions were employed in this study, with the first being the maximum absolute error ($MAE$) , defined as follows:

    \begin{table}[tbhp]
		\centering
		%\captionsetup{labelsep=period,labelfont=bf}
		\caption{Neural network training consistent hyper-parameters.}
		\label{Table:para}
		\begin{tabular}{*{6}{c}}
			\toprule
			Adam-iters&\qquad  L-BFGS-iters  &\qquad Adam-lr/L-BFGS-lr    &\qquad  $k$  \\
			\midrule
			
			$2000$  &\qquad $5000$ &\qquad  $0.001$  &\qquad $2$  \\
			% $40962$&    & &   & & &  \\
			%3.7841e-7}  & &   &  &  85.70\\
		\bottomrule
	    \end{tabular}
     \end{table}

    The assessment of prediction accuracy is carried out through the relative $L_2$ error between the reference solution $U(x)$ and the prediction solution $\widehat{U}(x)$,  
    
    \begin{align}\label{eq:18}
		\begin{split}
			L_2=\frac{\sqrt{\sum_{i=1}^N{\left| \widehat{U}\left( x_i,t_i \right) -U\left( x_i,t_i \right) \right|}^2}}{\sqrt{\sum_{i=1}^N{\left| U\left( x_i,t_i \right) \right|}^2}}. 
		\end{split}
    \end{align}    
   In the testing, a $100×100$ uniform grid is used to compute the relative $L_2$ error.

 %In all numerical experiments, the training dataset comprises 100 initial points ($|\mathcal{X}_i| = 100$), 200 boundary points ($|\mathcal{X}_b| = 200$), and 2000 collocation points ($|\mathcal{X}_p| = 2000$) uniformly distributed in the computational domain. This includes 1500 stationary collocation points ($|\mathcal{X}_{p,i}| = 1500$) and 500 adaptive collocation points ($|\mathcal{X}_{p,m}| = 500$). 

It is important to note that the L-BFGS optimizer can become trapped in local minima. To mitigate this issue, as it is mentioned in \cite{lu2021deepxde} we applies the Adam optimizer for a predetermined number of iterations at the beginning of each training phase. The use of Adam helps guide the parameters towards a more favorable initial state. After this initial phase, we use L-BFGS optimizer to expedite convergence.

\vspace{3em}

%\\ \hspace*{\fill} \\

\noindent \textbf{Example 1:  General Black-Scholes equation}

In this experiment, we are considering a general Black-Scholes equation. The 
test model is 
\begin{align}\label{eq:19}
	\begin{split}
		 \frac{\partial{U}}{\partial t} + \frac{1}{2}\sigma ^2S^2\frac{\partial^2{U}}{\partial S^2}+rS\frac{\partial{U}}{\partial S}-rU + M\left( S,t \right) = 0 , \ \ \  (S,t) \in \left( 0,1 \right) \times \left( 0,1 \right).\\
	\end{split}
\end{align}
Assume the exact solution is
\begin{align*}\label{eq:21}
	\begin{split}
		U\left( S,t \right) =\cos \left( 2\pi t \right) \cos \left( 2\pi S \right).
	\end{split}
\end{align*}
Then we can compute the corresponding function 
\begin{align*}
	\begin{split}
M\left( S,t \right) =&\left( r+2\pi ^2\sigma ^2S^2 \right) \cos \left( 2\pi S \right) \cos \left( 2\pi t \right)\\ 
		&+2\pi \cos \left( 2\pi S \right) \sin \left( 2\pi t \right) 
		+ 2\pi rS\sin \left( 2\pi S \right) \cos \left( 2\pi t \right).
\end{split}
\end{align*}
The following are the boundary and the terminal value conditions
\begin{align}
    \begin{split}
        \left\{
        \begin{array}{c}
            u\left( 0,t \right) = \cos \left( 2\pi t \right), \\
            u\left( 1,t \right) = \cos \left( 2\pi t \right), \\
            u\left( s,1 \right) = \cos \left( 2\pi s \right).
        \end{array}
        \right.
    \end{split}
\end{align}

% 导入图像阵列  
\begin{figure}[htbp]  
	\centering
	
	% 创建第一行第一列的子图  
	\begin{subfigure}{.19\textwidth}  
		\centering  
		\includegraphics[width=1.00\linewidth]{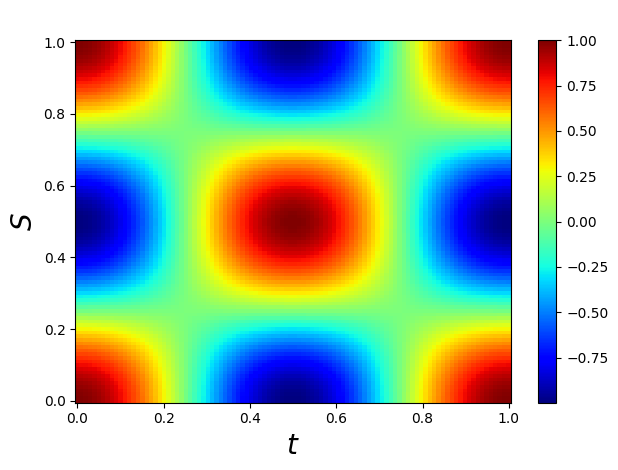}  
		\caption*{Exact}  
		\label{fig:11-1}  
	\end{subfigure}%  
	% 创建第一行第二列的子图  
	\begin{subfigure}{.19\textwidth}  
		\centering  
		\includegraphics[width=1.00\linewidth]{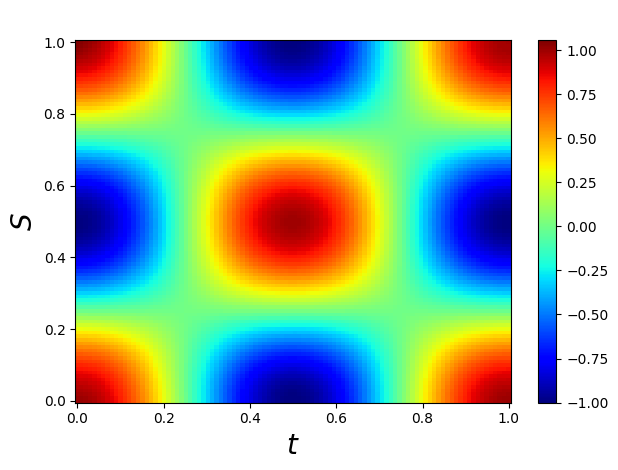}  
		\caption*{PINN}  
		\label{fig:11-2}  
	\end{subfigure}  
	% 创建第一行第三列的子图  
	\begin{subfigure}{.19\textwidth}  
		\centering  
		\includegraphics[width=1.00\linewidth]{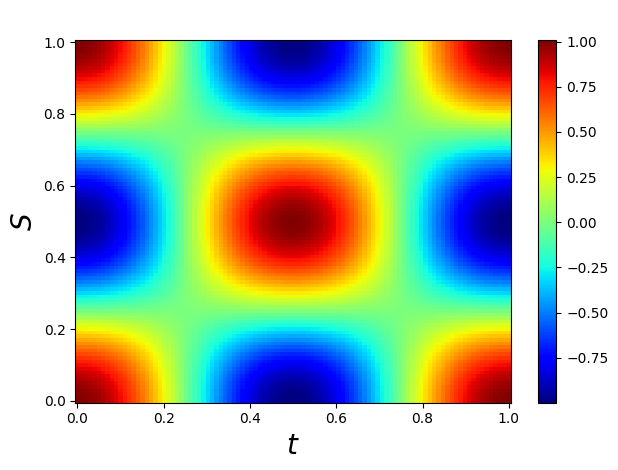}  
		\caption*{RAM-PINN}  
		\label{fig:11-3}  
	\end{subfigure}  
        \begin{subfigure}{.19\textwidth}  
		\centering  
		\includegraphics[width=1.00\linewidth]{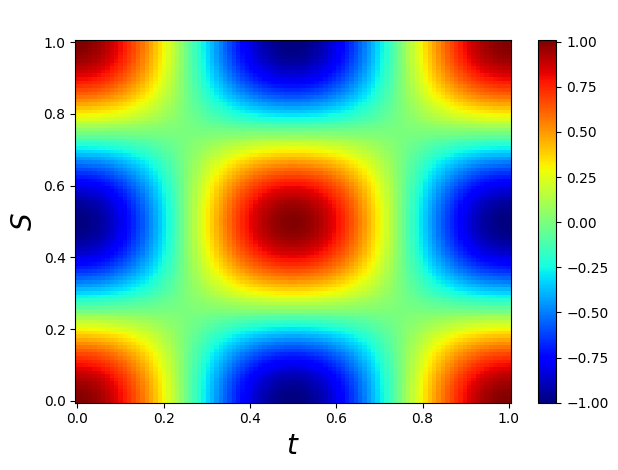}  
		\caption*{WAM-PINN}  
		\label{fig:11-4}
	\end{subfigure}  
		\begin{subfigure}{.19\textwidth}  
		\centering  
		\includegraphics[width=1.00\linewidth]{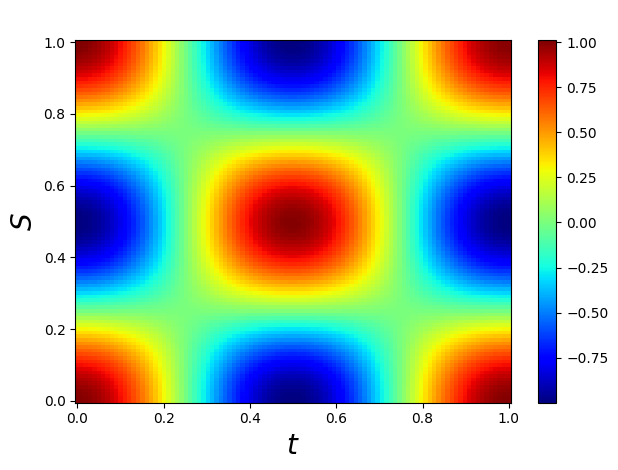}  
		\caption*{AM-PIRN}  
		\label{fig:11-5}
	\end{subfigure} 
	% 为整个图像阵列添加标题和标签  
	\caption{The exact solution and numerical solutions of PINN, RAM-PINN, WAM-PINN and AM-PIRN of the general Black-Scholes equation (\ref{eq:19}).} 
    % 整个图像阵列的标题  
	\label{fig11} % 整个图像阵列的标签  
	
\end{figure}

In this experiment, the asset volatility and interest rate are fixed at $\sigma = 0.25$ and $r = 0.05$, respectively. The training dataset consists of 100 initial points ($|\mathcal{X}_i| = 100$), 200 boundary points ($|\mathcal{X}_b| = 200$), and 2000 collocation points ($|\mathcal{X}_p| = 2000$) uniformly sampled across the computational domain. These collocation points include 1500 stationary points ($|\mathcal{X}_{p,i}| = 1500$) and 500 adaptively redistributed points ($|\mathcal{X}_{p,m}| = 500$).

\begin{figure} [!htp]  
	\centering  
	
	% 第一行
	\begin{subfigure}{.24\textwidth}  
		\centering  
		\includegraphics[width=1.00\linewidth]{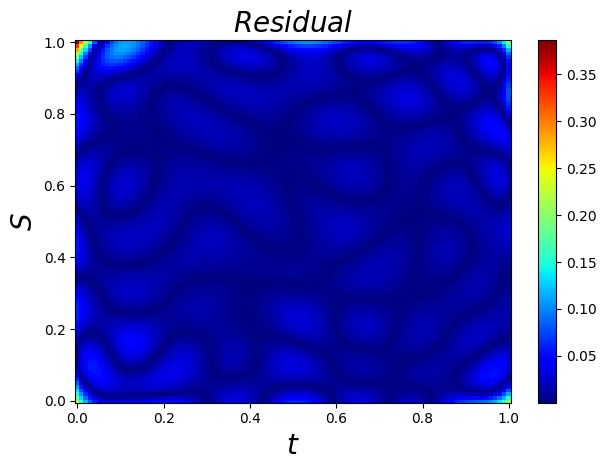}  
		%\caption*{PINN}  
		%\label{fig:14-1}  
	\end{subfigure}%  
	\begin{subfigure}{.24\textwidth}  
		\centering  
		\includegraphics[width=1.00\linewidth]{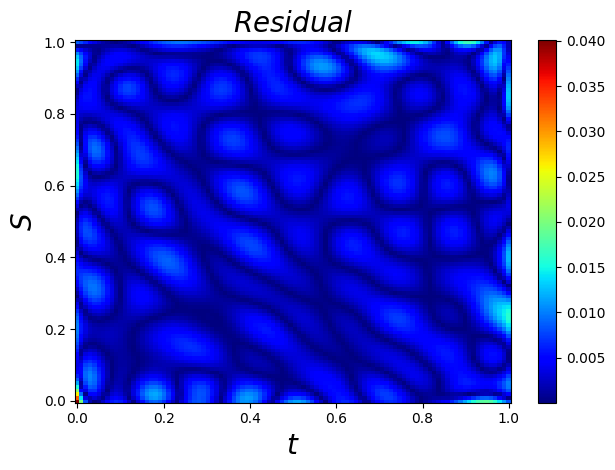}  
		%\caption*{RAR-PINN}  
		%\label{fig:14-2}  
	\end{subfigure}%  
	\begin{subfigure}{.24\textwidth}  
		\centering  
		\includegraphics[width=1.00\linewidth]{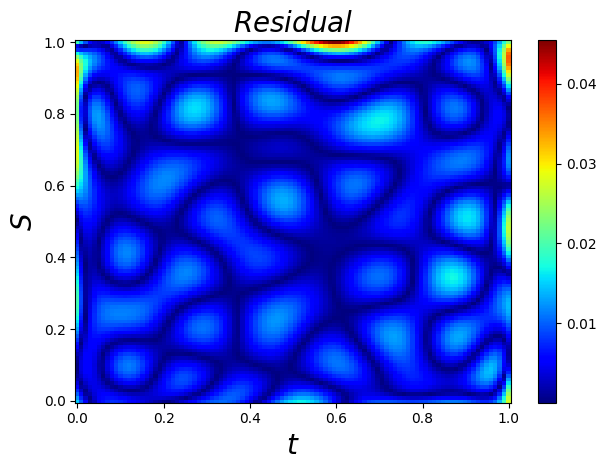}  
		%\caption*{RAR-PIRN}  
		%\label{fig:14-3}  
	\end{subfigure}%
        \begin{subfigure}{.24\textwidth}  
		\centering  
		\includegraphics[width=1.00\linewidth]{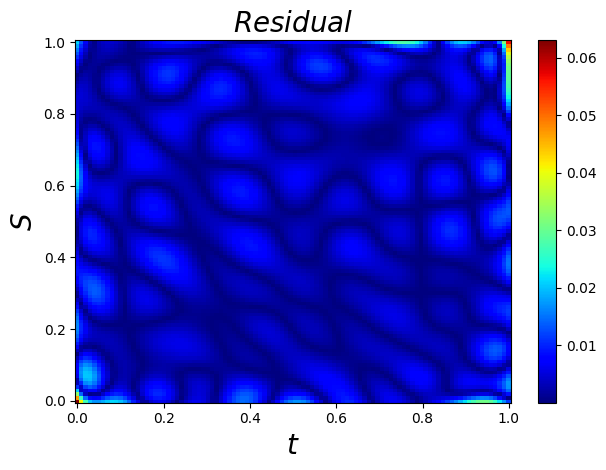}  
		%\caption*{RAR-PIRN}  
		%\label{fig:14-4}  
	\end{subfigure}%
	
	% 第二行
	\begin{subfigure}{.24\textwidth}  
		\centering  
		\includegraphics[width=1.00\linewidth]{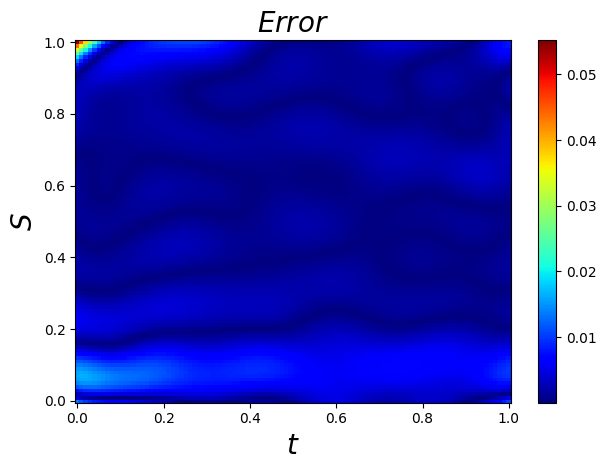}  
		%\caption*{PINN}  
		%\label{fig:14-4}  
	\end{subfigure}%  
	\begin{subfigure}{.24\textwidth}  
		\centering  
		\includegraphics[width=1.00\linewidth]{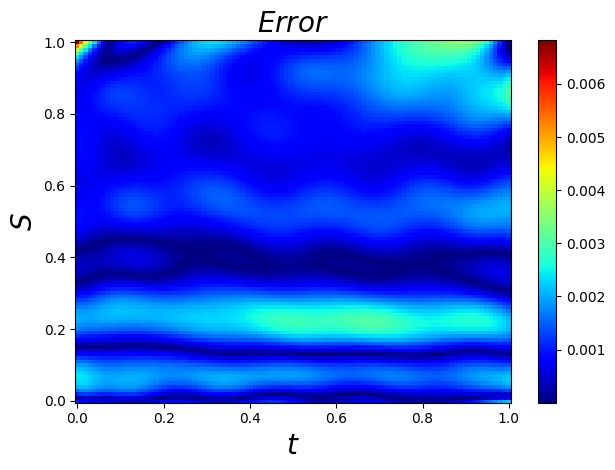}  
		%\caption*{RAR-PINN}  
		%\label{fig:14-5}  
	\end{subfigure}%  
	\begin{subfigure}{.24\textwidth}  
		\centering  
		\includegraphics[width=1.00\linewidth]{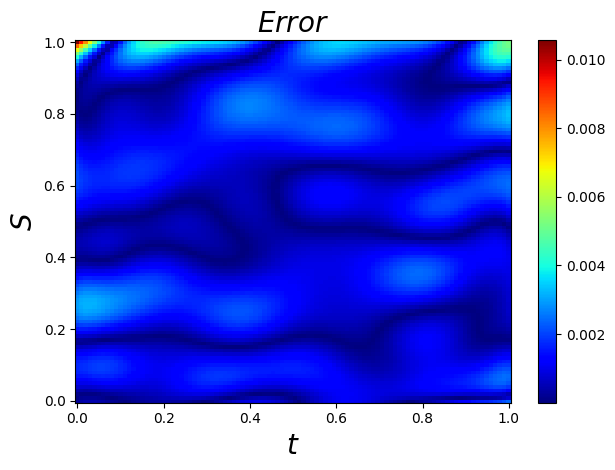}  
		%\caption*{RAR-PIRN}  
		%\label{fig:14-6}  
	\end{subfigure}%
		\begin{subfigure}{.24\textwidth}  
		\centering  
		\includegraphics[width=1.00\linewidth]{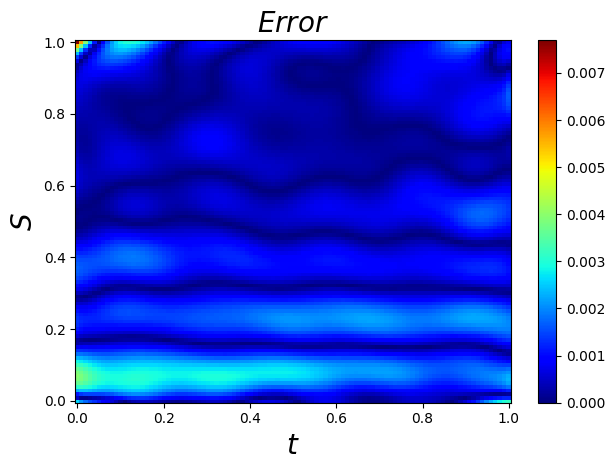}  
		%\caption*{RAR-PIRN}  
		%\label{fig:14-6}  
	\end{subfigure}%
    
	% 第三行
	\begin{subfigure}{.24\textwidth}  
		\centering  
		\includegraphics[width=1.00\linewidth]{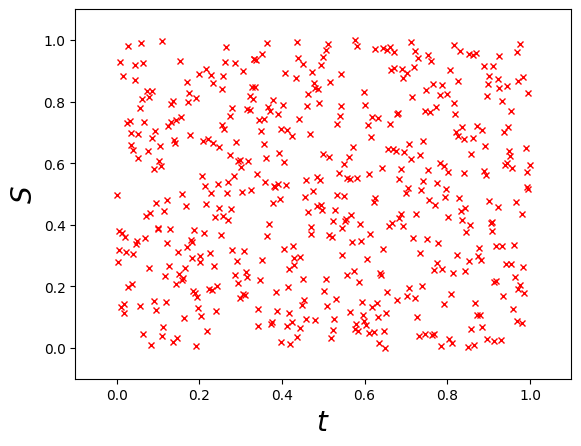}  
		\caption*{PINN}  
		%\label{fig:14-7}  
	\end{subfigure}%  
	\begin{subfigure}{.24\textwidth}  
		\centering  
		\includegraphics[width=1.00\linewidth]{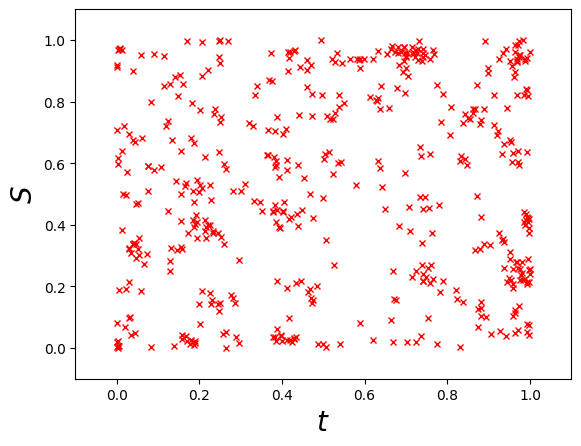}  
		\caption*{RAM-PINN}  
		%\label{fig:14-8}  
	\end{subfigure}%  
        \begin{subfigure}{.24\textwidth}  
		\centering  
		\includegraphics[width=1.00\linewidth]{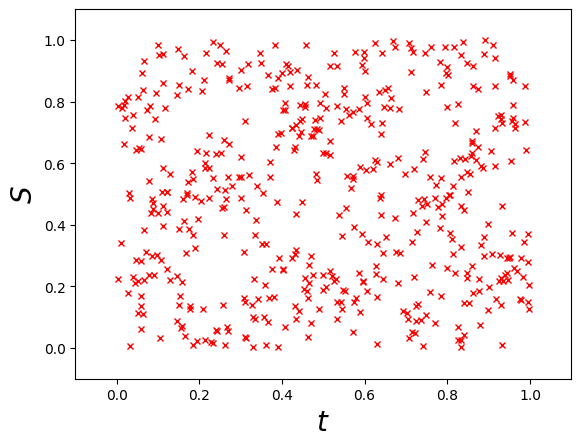}  
		\caption*{WAM-PINN}  
		%\label{fig:14-9}  
	\end{subfigure}%
	\begin{subfigure}{.24\textwidth}  
		\centering  
		\includegraphics[width=1.00\linewidth]{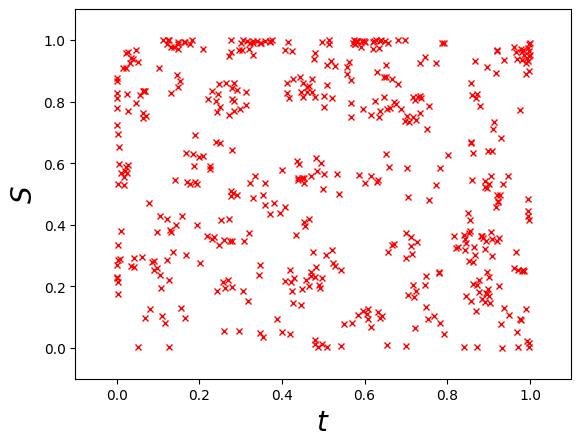}  
		\caption*{AM-PIRN}  
		%\label{fig:14-9}  
	\end{subfigure}%
		% 为整个图像阵列添加标题和标签  
	\caption{The PDE residuals (first row), absolute errors of estimated solutions (middle row), and distribution of mobile collocation points (last row) for the general Black-Scholes equation (\ref{eq:19}) computed by PINN, RAM-PINN, WAM-PINN and AM-PIRN after $10$ iteration rounds, respectively.}
	\label{fig14} % 整个图像阵列的标签
\end{figure}

Fig. \ref{fig11} illustrates the exact solution against numerical approximations from PINN, RAM-PINN, WAM-PINN, and AM-PIRN. Under identical training conditions, Fig.  \ref{fig14} demonstrates that AM-PIRN achieves a significantly lower PDE residual compared to PINN. Quantitatively, the maximum residuals for RAM-PINN, AM-PIRN, and WAM-PINN are $0.04$, $0.07$, and $0.08$, respectively, with corresponding maximum absolute errors of 0.035, 0.02, and 0.03—all markedly lower than PINN’s residual (0.6) and error (0.08). Notably, while the residuals differ substantially across methods, the absolute errors in option price approximations remain comparable (e.g., AM-PIRN’s error is 0.02 and RAM-PINN’s 0.035). This underscores a critical insight:  reducing PDE residuals does not guarantee a proportional improvement in pricing accuracy, highlighting the nuanced relationship between residual minimization and functional error control. 

The final row of Fig. \ref{fig14} illustrates the collocation point distributions for all methods. Notably, the movable collocation points $\mathcal{X}_{p,m}$ 
in WAM-PINN exhibit a nearly uniform distribution. In contrast, $\mathcal{X}_{p,m}$ 
for RAM-PINN and AM-PIRN are densely clustered near the domain boundary $S=1$. This concentration aligns with the expectation that regions near boundaries often experience larger residuals and errors, making such adaptive sampling strategies computationally advantageous. 
%In the last row of Fig. \ref{fig14},  we provide the collocation point distribution for all the methods. It reveals that the movable collocation points   for WAM-PINN are almost uniform. In contrast, $\mathcal{X}_{p,m}$ for RAM-PINN and AM-PIRN are  distributed densely near the boundary domain $S=1$. It is reasonable since the large residuals and errors of functions usually happened near the boundaries.  %It reveals that the movable collocation points $\mathcal{X}_{p,m}$  for AM-PINN is mainly concentrated near $t=0$, where the higher residuals appear.  It means that in AM-PINN the error of the equation accumulates over time, with the maximum error occurring at the initial moment. In contrast, $\mathcal{X}_{p,m}$ for AM-PIRN is  distributed within  all the interior domain. 

\begin{figure} [!htp]  
	\centering  
		\includegraphics[width=0.6\linewidth]{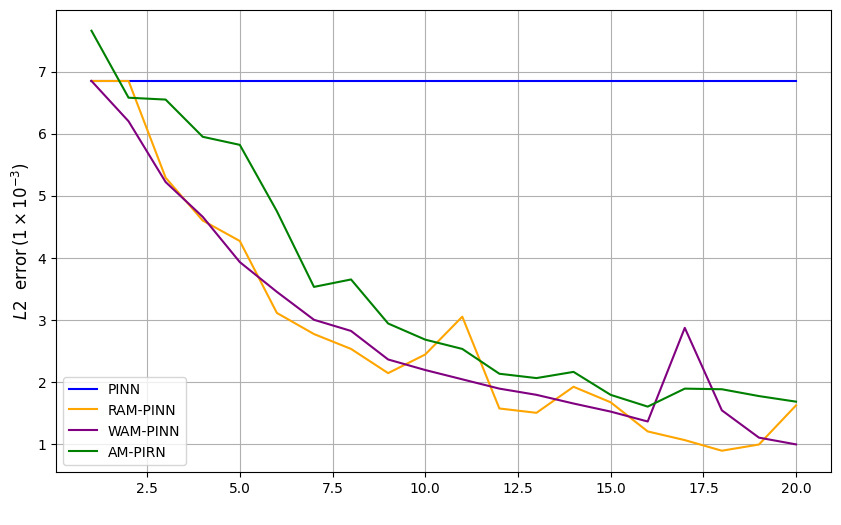}  
    	% 为整个图像阵列添加标题和标签  
	\caption{The iterative convergence curves of the $L_2$ errors of the estimated solutions for the general Black-Scholes equation by PINN, RAM-PINN, WAM-PINN and AM-PIRN.} % 整个图像阵列的标题  
	\label{fig:BS_convergence} % 整个图像阵列的标签
\end{figure}

Fig. \ref{fig:BS_convergence} demonstrates the convergence behavior of relative $L_2$ errors for RAM-PINN, WAM-PINN, and AM-PIRN over 20 adaptive iteration rounds. As the number of adaptive collocation rounds increases, all three methods achieve a final error reduction to the order of $10^{-2}$, validating their asymptotic convergence. However, RAM-PINN and WAM-PINN exhibit significantly greater oscillations in their error trajectories compared to the monotonic decrease observed in AM-PIRN. This contrast highlights AM-PIRN’s superior stability during training, suggesting its enhanced robustness in handling boundary-sensitive dynamics inherent to the Black-Scholes equation. The results align with the collocation point distributions in Fig. \ref{fig14}, where AM-PIRN’s smoother convergence corresponds to its strategic error minimization near critical regions like $S=1$. 
 
We further investigate the impact of collocation point allocation by conducting experiments with varying numbers of movable ($\mathcal{X}_{p,m}$) and immovable ($\mathcal{X}_{p,i}$) points. In Table (\ref{Tab:BS1}), we fix 
$|\mathcal{X}_{p,m}| = 500$ and vary $\mathcal{X}_{p,i}$ from 1000 to 2000. Adaptive methods consistently outperform PINN: for $|\mathcal{X}_{p,i}| = 1500$, AM-PIRN achieves a PDE residual of $7.07\times 10^{-5}$ and a relative $L_2$ error of $2.68\times 10^{-3}$, both an order of magnitude lower than PINN’s $3.71\times 10^{-4}$ (residual) and $7.06\times 10^{-3}$ (error). Similarly, in Table (\ref{Tab:BS2}), we fix $|\mathcal{X}_{p,i}| = 1500$ and vary $|\mathcal{X}_{p,m}|$ from 250 to 2000. Here, the adaptive sampling methods attains the lower error (about $2\times 10^{-3}$) at $|\mathcal{X}_{p,m}| = 500$, outperforming PINN ($7\times 10^{-3}$).  Notably, AM-PIRN exhibits stable performance across configurations, with residuals and errors consistently below $1.54\times 10^{-4}$ and $4.68\times 10^{-3}$, respectively. These results confirm that adaptive sampling strategies significantly enhance accuracy compared to PINN frameworks.

 %the function maximum absolute errors and relative $L_2$ errors  of the equations of AM-PIRN are relatively smaller than PINN ,RAM-PINN, WAM-PINN. This further demonstrates the effectiveness of AM-PIRN, as it ensures function approximation accuracy while minimizing equation residuals. 

% 导入图像阵列  
% 导入图像阵列  

\begin{table}[!htp]
    \centering
\begin{minipage}[c]{0.49\textwidth}
\centering
\footnotesize
\begin{tabular}{@{}lccc@{}}
\toprule
$|\mathcal{X}_{p,i}|$ & 1000 & 1500 & 2000 \\ \midrule
PINN & 8.83E$-$04 & 3.71E$-$04 & 7.04E$-$04 \\
RAM-PINN & 3.65E$-$05 & 2.72E$-$05 & 5.43E$-$05 \\
WAM-PINN & 2.60E$-$05 & 3.61E$-$05 & 4.41E$-$05 \\
AM-PIRN & 1.53E$-$04 & 7.07E$-$05 & 1.54E$-$04 \\
\bottomrule
\end{tabular}
%\captionof{table}{The residuals of the general Black-Scholes equation with $|\mathcal{X}_{p,m}| = 500$ movable points.  \label{Tab:BS1}}
\end{minipage}
\begin{minipage}[c]{0.49\textwidth}
\centering
\footnotesize
\begin{tabular}{@{}lccc@{}}
\toprule
$|\mathcal{X}_{p,i}|$ & 1000 & 1500 & 2000 \\ \midrule
PINN & 1.27E$-$02 & 7.06E$-$03 & 9.29E$-$03 \\
RAM-PINN & 2.59E$-$03 & 1.91E$-$03 & 1.65E$-$03 \\
WAM-PINN & 2.22E$-$03 & 2.10E$-$03 & 1.92E$-$03 \\
AM-PIRN & 3.34E$-$03 & 2.68E$-$03 & 4.68E$-$03 \\
\bottomrule
\end{tabular}
%\captionof{table}{The relative $L_2$-errors of the general Black-Scholes equation with $|\mathcal{X}_{p,m}| = 500$ movable points. \label{Tab:BS2}}
\end{minipage}
 \caption{The PDE residuals (left) and the relative $L_2$ errors (right) of the general Black-Scholes equation with $|\mathcal{X}_{p,m}| = 500$ movable points, respectively.  }
    \label{Tab:BS1}
\end{table}

%\\ \hspace*{\fill} \\

\begin{table}[!htp]
    \centering
\begin{minipage}[c]{0.49\textwidth}
\centering
\footnotesize
\begin{tabular}{@{}lccc@{}}
\toprule
$|\mathcal{X}_{p,m}|$ & 250 & 500 & 750 \\ \midrule
PINN & 4.98E$-$04 & 3.71E$-$04 & 3.25E$-$04 \\
RAM-PINN & 5.09E$-$05 & 2.72E$-$05 & 2.88E$-$05 \\
WAM-PINN & 4.42E$-$05 & 3.61E$-$05 & 2.01E$-$05 \\
AM-PIRN & 1.57E$-$04 & 7.07E$-$05 & 2.94E$-$04 \\
\bottomrule
\end{tabular}
%\captionof{table}{The residuals of the general Black-Scholes equation with $|\mathcal{X}_{p,i}| = 1500$ stationary points.  \label{Tab:BS3}}
\end{minipage}
\begin{minipage}[c]{0.49\textwidth}
\centering
\footnotesize
\begin{tabular}{@{}lccc@{}}
\toprule
$|\mathcal{X}_{p,m}|$ & 250 & 500 & 750 \\ \midrule
PINN & 8.86E$-$03 & 7.06E$-$03 & 7.25E$-$03 \\
RAM-PINN & 3.26E$-$03 & 1.91E$-$03 & 3.69E$-$03 \\
WAM-PINN & 1.86E$-$03 & 2.10E$-$03 & 1.94E$-$03 \\
AM-PIRN & 3.91E$-$03 & 2.68E$-$03 & 4.57E$-$03 \\
\bottomrule
\end{tabular}
%\captionof{table}{The relative $L_2$-errors of the general Black-Scholes equation with $|\mathcal{X}_{pi}| = 1500$ stationary points. \label{Tab:BS4} }
\end{minipage}
 \caption{The PDE residuals (left) and the relative $L_2$ errors (right) of the general Black-Scholes equation with $|\mathcal{X}_{p, i}| = 1500$ stationary points, respectively.}
    \label{Tab:BS2}
\end{table}

\vspace{3em}

%\\ \hspace*{\fill} \\

\noindent \textbf{Example 2: Barles' and Soner's model}

In this example, we compare the  aforementioned methods to solve the nonlinear Black-Scholes equation (\ref{BS}) with the volatility being determined by Barles' and Soner's model in  \cite{hodges1989optimal}. Here the nonlinear volatility is given by 
	
	\begin{align}\label{eq:3}
		\begin{split}
			\sigma^2(S,t)=\sigma _{0}^{2}\left( 1+\varPsi \left( e^{r\left( T-t \right)}a^2S^2 \frac{\partial^2 U}{\partial S^2} \right) \right) ,
		\end{split}
	\end{align}
	where $\sigma_0$ is the asset volatility, and $a$ is transaction cost. $\varPsi \left( x \right) $ denotes the solution of the following nonlinear ordinary differential equation
	\begin{equation*}\label{eq:4}
		\begin{split}
			\varPsi ^{'}\left( x \right) =\frac{\varPsi \left( x \right) +1}{2\sqrt{x\varPsi \left( x \right)}-x}\,\,,  x\ne 0 ,
		\end{split}
	\end{equation*}
	with the initial condition $\varPsi \left( 0 \right) =0$.
The analysis of this ordinary differential equation by Barles and Soner in \cite{barles1998option} demonstrates that
	\begin{align*}\label{eq:6}
		\begin{split}
			\underset{x\rightarrow \infty}{\lim}\frac{\varPsi \left( x \right)}{x}=1     \; \text{and}\;  \underset{x\rightarrow -\infty}{\lim}\frac{\varPsi \left( x \right)}{x}=-1.
		\end{split}
	\end{align*}
This property allows treating the function $\varPsi \left( \cdot \right)$ as the identity for large arguments, leading to the transformation of volatility	
	\begin{equation}\label{eq:7}
		\begin{split}
			\sigma^2=\sigma _{0}^{2}\left( 1+e^{r\left( T-t \right)}a^2S^2 \frac{\partial^2 U}{\partial S^2} \right).
		\end{split}
	\end{equation}

In this example, we set the maturity  $T=1$, and the asset $S$ being located in $ \left[ 0,80 \right]$. The so-called Butterfly Spread option is tested, which has specific initial conditions, 
\begin{align*}\label{eq:22}
	\begin{split}
	U\left( S,T\right) =\max \left( S-K_1,T\right) -2\max \left( S-K_2,T \right) +\max \left( S-K_3,T\right),
	\end{split}
\end{align*}
with $K_1=30$, $K_2=40$, and $K_3=50$, and the boundary conditions
\begin{equation*}  \label{eq:23}
	\begin{split}
		U\left( 0,t \right) =U\left( 80,t \right) =0. 
	\end{split}
\end{equation*}
The other parameters used for this experiment are given by 
		$r=0.1,a=0.02$, and $\sigma _0=0.2$.

% 导入图像阵列  
\begin{figure}[htbp]  
	\centering  
	% 创建第一行第一列的子图  
	\begin{subfigure}{.24\textwidth}  
		\centering  
		\includegraphics[width=.99\linewidth]{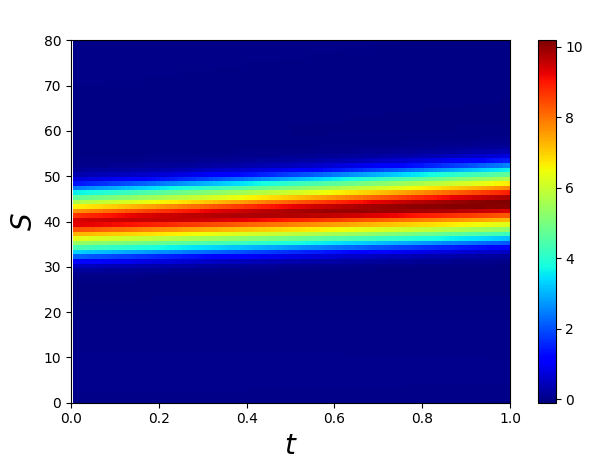}  
        \caption*{PINN}  
		%\label{fig:16-1}  
	\end{subfigure}%  
	% 创建第一行第二列的子图  
	\begin{subfigure}{.24\textwidth}  
		\centering  
		\includegraphics[width=.99\linewidth]{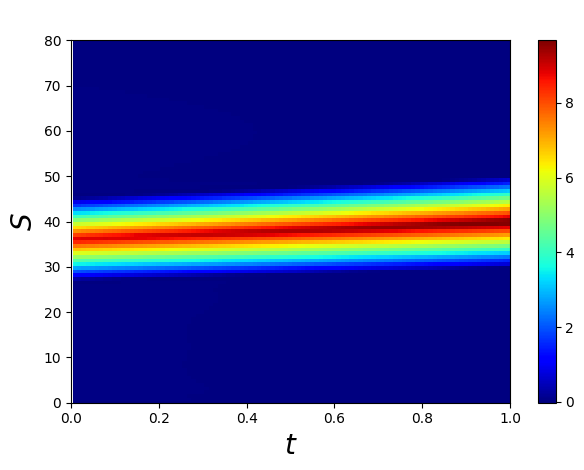} 
        \caption*{RAM-PINN}  
		%\label{fig:16-2}  
	\end{subfigure}%  
        % 创建第一行第四列的子图  
	\begin{subfigure}{.24\textwidth}  
		\centering  
		\includegraphics[width=.99\linewidth]{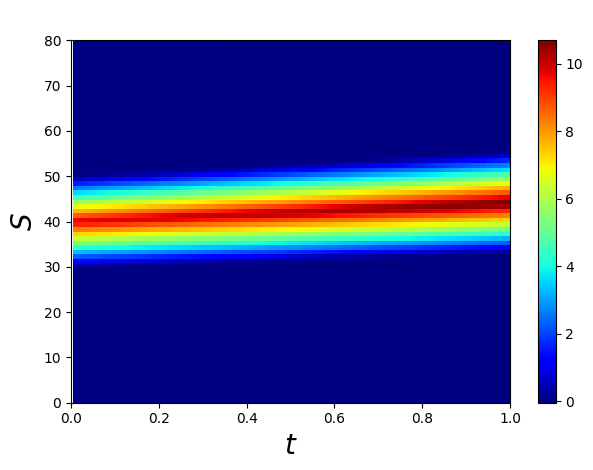}
        \caption*{WAM-PINN}  
		%\label{fig:16-3}  
	\end{subfigure}% 
        % 创建第一行第三列的子图  
	\begin{subfigure}{.24\textwidth}  
		\centering  
		\includegraphics[width=.99\linewidth]{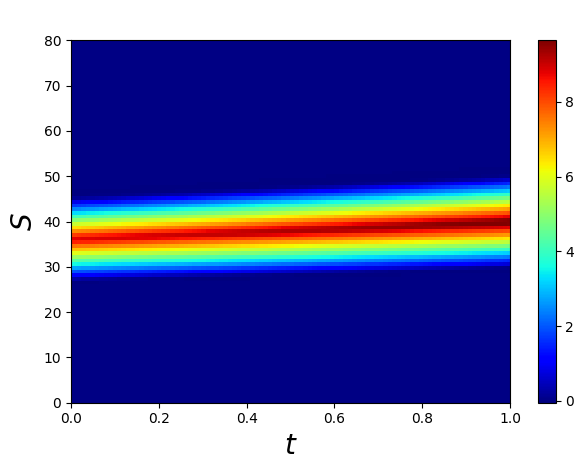}
        \caption*{AM-PIRN}  
		%\label{fig:16-3}  
	\end{subfigure}%

	% 为整个图像阵列添加标题和标签  
	\caption{
    The numerical solutions of Barles’ and Soner’s model given by PINN, RAM-PINN, WAM-PINN and AM-PIRN, respectively.} % 整个图像阵列的标题  
	\label{fig16} % 整个图像阵列的标签
\end{figure}

%Consequently, we have generated 3D plots of the numerical solution and residual plots to visually demonstrate the model's approximation effect, as illustrated in Fig.\ref{fig16} and Fig.\ref{fig17}.

% 导入图像阵列  
\begin{figure}  
	\centering  
	% 创建第一行第一列的子图  
	\begin{subfigure}{.24\textwidth}  
		\centering  
		\includegraphics[width=.9\linewidth]{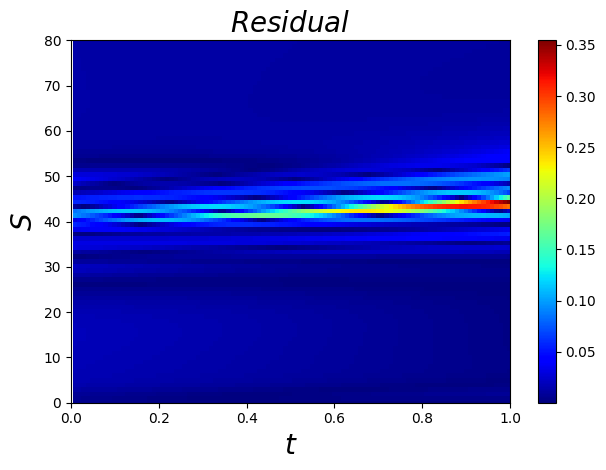}  
		\label{fig:17-1}  
	\end{subfigure}%  
        % 创建第一行第二列的子图  
	\begin{subfigure}{.24\textwidth}  
		\centering  
		\includegraphics[width=.9\linewidth]{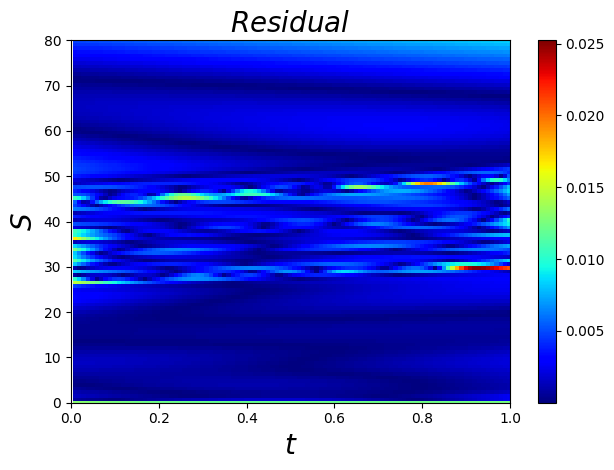}  
		\label{fig:17-2}  
	\end{subfigure}% 
        % 创建第一行第四列的子图  
	\begin{subfigure}{.24\textwidth}  
		\centering  
		\includegraphics[width=.9\linewidth]{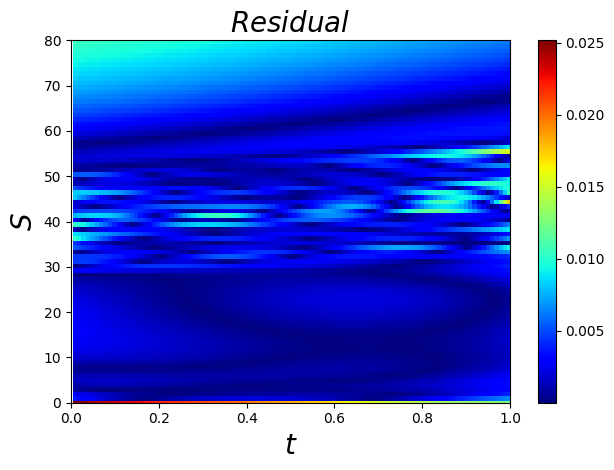}  
		\label{fig:17-4}  
	\end{subfigure}%  
        % 创建第一行第三列的子图  
	\begin{subfigure}{.24\textwidth}  
		\centering  
		\includegraphics[width=.9\linewidth]{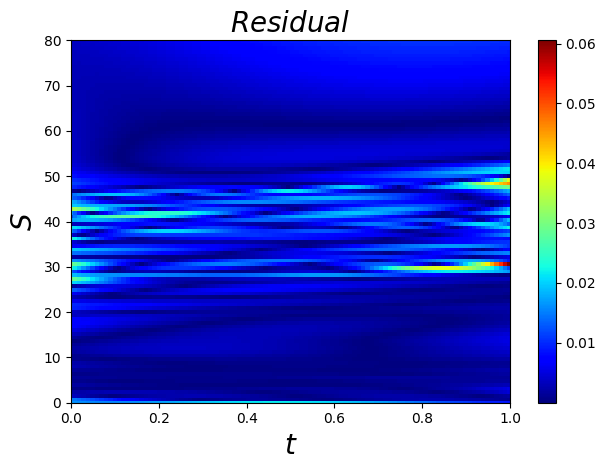}  
		\label{fig:17-3}  
	\end{subfigure}%  

	% 创建第一行第二列的子图  
	\begin{subfigure}{.24\textwidth}  
		\centering  
		\includegraphics[width=0.9\linewidth]{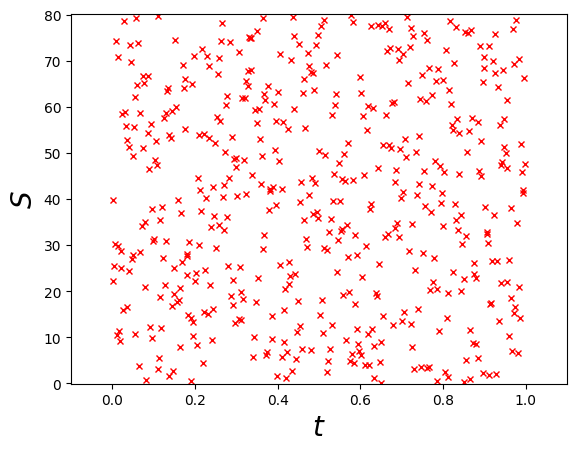}  
        \caption*{PINN}  
		\label{fig:17-5}  
	\end{subfigure} 
        % 创建第一行第二列的子图  
	\begin{subfigure}{.24\textwidth}  
		\centering  
		\includegraphics[width=0.9\linewidth]{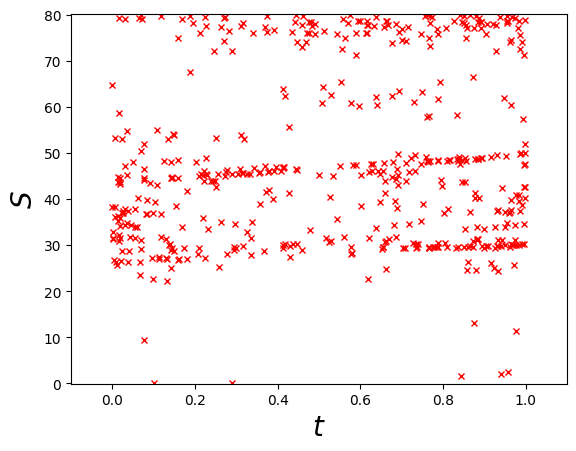} 
        \caption*{RAM-PINN}  
		\label{fig:17-6}  
	\end{subfigure} 
        % 创建第一行第二列的子图  
	\begin{subfigure}{.24\textwidth}  
		\centering  
		\includegraphics[width=.9\linewidth]{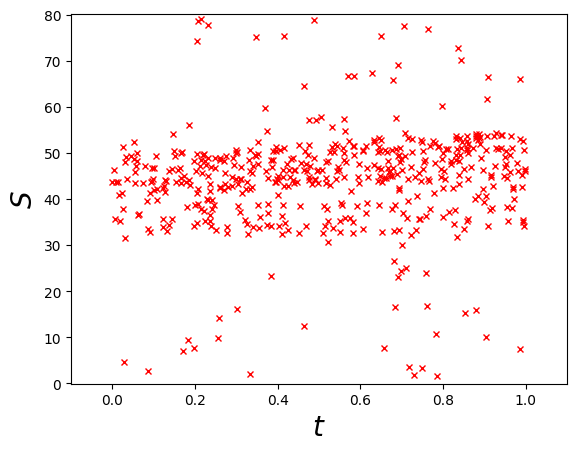} 
        \caption*{WAM-PINN}  
		\label{fig:17-8}  
	\end{subfigure} 
        % 创建第一行第二列的子图  
	\begin{subfigure}{.24\textwidth}  
		\centering  
		\includegraphics[width=.9\linewidth]{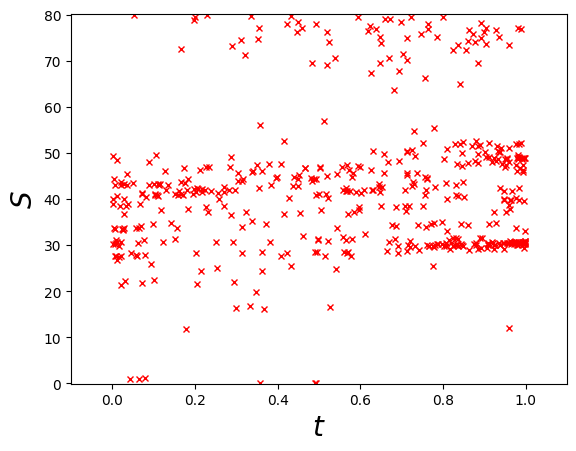} 
        \caption*{AM-PIRN}  
		\label{fig:17-7}  
	\end{subfigure} 

	% 为整个图像阵列添加标题和标签  
	\caption{
The PDE residuals (first row) and distribution of mobile collocation points (second row) for  Barles’ and Soner’s model computed by PINN, RAM-PINN, WAM-PINN and AM-PIRN after $10$ iteration rounds, respectively.}
	\label{fig17} % 整个图像阵列的标签
\end{figure}

Given the absence of a closed-form solution for Barles’ and Soner’s nonlinear volatility model, we compute the numerical solutions using PINN, RAM-PINN, WAM-PINN, and AM-PIRN, with collocation point allocations identical to Example 1: fixed initial/boundary points $|\mathcal{X}_i| = 100$ and $|\mathcal{X}_b| = 200$, along with $|\mathcal{X}_{p,i}| = 1500$ fixed and $|\mathcal{X}_{p,m}| = 500$ movable collocation points. As shown in Fig.~\ref{fig16}, all methods successfully resolve the sharp gradient features near the strike price range $S \in [30, 50]$, where the option price exhibits its steepest spatial variation. This numerical behavior aligns precisely with the theoretical expectations of the model, which predicts localized nonlinear volatility effects in this region. The consistency across adaptive and non-adaptive methods underscores the robustness of the collocation framework in capturing boundary-layer dynamics inherent to such financial PDEs.

Fig.~\ref{fig17} illustrates the PDE residual distributions (top row) and movable collocation point patterns (bottom row) for Barles' and Soner's nonlinear volatility model after 10 iterations. The PINN residual figure exhibits maximum values up to 0.35 concentrated in the high-gradient zone $ S \in [30, 50]$ with $t \in [0.4, 1.0]$, directly corresponding to the steepest price variations in the solution profile. Adaptive methods demonstrate significantly reduced residuals, with RAM-PINN and AM-PIRN showing peak values below 0.06 through strategic collocation point allocation. Their movable points form a dense band between $S=30$ and $S=50$, mirroring the spatial coordinates of maximum solution curvature. Notably, a few number of RAM-PINN's and AM-PIRN's adaptive points cluster near the boundary $S=80$, addressing secondary error accumulation observed at $t=0.8$ with residual magnitudes below 0.04.

%As depicted in Fig. \ref{fig17}, the region with large residuals of PINN is coincident with the function steep region.  In all the adaptive methods, there is a distinct banded area from $S=30$ to $S=50$ with the clustered collocation points.  The adaptive movement points of RAM-PINN and AM-PIRN are gathered also near the boundary $S=80$. 

%The adaptive movement points of RAM-PINN are gathered near the strike price $K_2=40$ and the boundary $S=80$, while in AM-PIRN there is a distinct banded area from $S=30$ to $S=50$ with the clustered collocation points.   

The PDE residuals for Barles' and Soner's model under varying collocation points are detailed in Table~(\ref{Tab:Barles}). With $|\mathcal{X}_{p,m}| = 500$ movable  points (left), adaptive methods consistently outperform PINN by 1--2 orders of magnitude. PINN exhibits residuals of about $10^{-3}$ across configurations, while WAM-PINN and RAM-PINN achieves about  $10^{-5}$. For variable fixed points $|\mathcal{X}_{p,i}| = 1500$ (right), AM-PIRN shows moderate performance with residuals between $3.64 \times 10^{-4}$ and $5.09 \times 10^{-4}$, still surpassing PINN by an order of magnitude. %Notably, increasing movable points to $|\mathcal{X}_{p,m}| = 750$ reduces WAM-PINN's residual by 34\% compared to its $|\mathcal{X}_{p,m}| = 250$ configuration ($3.68 \times 10^{-5}$ vs. $2.63 \times 10^{-5}$), highlighting the value of adaptive allocation. 
These results confirm that dynamic collocation strategies significantly enhance numerical precision in nonlinear volatility modeling.

%As the former example,  we then test more experiments  with different movable and immovable collocation points in Table (\ref{Tab:Barles}).  The residuals  of the equations of  all the adaptive metheods  are  one or two orders smaller than PINN.

\begin{comment}
\begin{table}[h]
\centering
\begin{tabular}{@{}lccc@{}}
\toprule
$|\mathcal{X}_{p,i}|$ & 1000 & 1500 & 2000 \\ \midrule
PINN & 5.91E$-$03 & 6.27E$-$03 & 6.11E$-$03 \\
RAM-PINN & 4.44E$-$05 & 4.47E$-$05 & 3.44E$-$05 \\
WAM-PINN & 4.43E$-$05 & 3.38E$-$05 & 3.68E$-$05 \\ 
AM-PIRN & 3.14E$-$04 & 4.52E$-$04 & 4.08E$-$04 \\ 
\bottomrule
\end{tabular}
\caption{The residuals of the Barles’ and Soner’s model with $|\mathcal{X}_{p,m}| = 500$ movable points.}
\label{Tab:EX7} 
\end{table}

\begin{table}[h]
\centering
\begin{tabular}{@{}lccc@{}}
\toprule
$|\mathcal{X}_{p,i}|$ & 250 & 500 & 750 \\ \midrule
PINN & 6.65E$-$03 & 6.27E$-$03 & 5.85E$-$03 \\
RAM-PINN & 5.38E$-$05 & 4.47E$-$05 & 3.38E$-$05 \\
WAM-PINN & 4.43E$-$05 & 3.38E$-$05 & 2.63E$-$05 \\ 
AM-PIRN & 5.09E$-$04 & 4.52E$-$04 & 3.64E$-$04 \\ 
\bottomrule
\end{tabular}
\caption{The  relative $L_2$-errors the Barles’ and Soner’s model with $|\mathcal{X}_{p,m}| = 1500$ stationary points.  }
\label{Tab:EX8} 
\end{table}
\end{comment}

\begin{table}[!htp]
    \centering
\begin{minipage}[c]{0.49\textwidth}
\centering
\footnotesize
\begin{tabular}{@{}lccc@{}}
\toprule
$|\mathcal{X}_{p,i}|$ & 1000 & 1500 & 2000 \\ \midrule
PINN & 5.91E$-$03 & 6.27E$-$03 & 6.11E$-$03 \\
RAM-PINN & 4.44E$-$05 & 4.47E$-$05 & 3.44E$-$05 \\
WAM-PINN & 4.43E$-$05 & 3.38E$-$05 & 3.68E$-$05 \\ 
AM-PIRN & 3.14E$-$04 & 4.52E$-$04 & 4.08E$-$04 \\ 
\bottomrule
\end{tabular}
%\captionof{table}{The residuals of the Barles’ and Soner’s model with $|\mathcal{X}_{p,m}| = 500$ movable points.\label{Tab:Barles_1} }
\end{minipage}
\begin{minipage}[c]{0.49\textwidth}
\centering
\footnotesize
\begin{tabular}{@{}lccc@{}}
\toprule
$|\mathcal{X}_{p,i}|$ & 250 & 500 & 750 \\ \midrule
PINN & 6.65E$-$03 & 6.27E$-$03 & 5.85E$-$03 \\
RAM-PINN & 5.38E$-$05 & 4.47E$-$05 & 3.38E$-$05 \\
WAM-PINN & 4.43E$-$05 & 3.38E$-$05 & 2.63E$-$05 \\ 
AM-PIRN & 5.09E$-$04 & 4.52E$-$04 & 3.64E$-$04 \\ 
\bottomrule
\end{tabular}
%\captionof{table}{The residuals of the Barles’ and Soner’s model with $|\mathcal{X}_{p,m}| = 1500$ stationary points. \label{Tab:Barles_2} }
\end{minipage}
\caption{The PDE residuals  of the Barles’ and Soner’s model  with 
$|\mathcal{X}_{p,m}| = 500$ movable points (left)  and  $|\mathcal{X}_{p, i}| = 1500$ (right) stationary points, respectively.}\label{Tab:Barles}
\end{table}

%the plots showcase the aggregation of residuals and resampling points for the RAR-PIRN solution of the nonlinear Black-Scholes equation. All the points cluster around $S=30$ to $S=50$, with the maximum value of residuals reaching only $5$. In the residual plots, residuals are close to 0 in most regions. This observation signifies the effectiveness of the RAR-PIRN approximation for the nonlinear Black-Scholes equation.

\vspace{3em}

%\\ \hspace*{\fill} \\

\noindent \textbf{Example 3: CEV model}

In this experiment, we will investigate the CEV model for European put option, in which the underlying asset is 
\begin{align*}
&dS_t = rS_t dt + \sigma S_t^{\beta/2} dW_t,
\end{align*}
and the option pricing equation is 
\begin{align*}
\begin{cases}
    \frac{\partial U}{\partial t} +\frac{1}{2} \sigma^2 S^\beta \frac{\partial^2 U}{\partial S^2} + rS \frac{\partial U}{\partial S} - rU =0,\\
 U(S,T)=(K-S)^+.
\end{cases}
\end{align*}

The solution for European put option \( U(S_t, t, K, r, \sigma) \) is
\[U(S_t, t, K, r, \sigma) = 
    \begin{cases} 
    Ke^{-r(T-t)}Q\left(2x; \frac{2}{2-\beta}, 2y\right) - S_t\left[1-Q\left(2y; 2 + \frac{2}{2-\beta}, 2x\right)\right], & \text{for } \beta < 2, \\
    Ke^{-r(T-t)}Q\left(2y; 2 + \frac{2}{\beta-2}, 2x\right) - S_t\left[1-Q\left(2x; \frac{2}{\beta-2}, 2y\right)\right], & \text{for } \beta > 2,
    \end{cases}\]
%\[
%    Q(a; b, c) = \int_{-\infty}^a \frac{1}{\sqrt{2\pi}} e^{-\frac{(t-b)^2}{2c^2}} dt
%\]
where \( Q(w;\nu, \lambda) \) is the complementary distribution function of a non-central chi-square law, \( \nu \) is the degree of freedom, \( \lambda \) is a non-centrality parameter, and
\begin{align*}
x &= S_t^{2-\beta} e^{r(2-\beta)(T-t)} d, & y &= K^{2-\beta} d, \\
d &= \frac{2 r}{\delta^2 (2-\beta) \left[ e^{r(2-\beta)(T-t)} - 1 \right]}, & \delta^2 &= \sigma^2 S_0^{2-\beta}.
\end{align*}

\begin{comment}
    
	\begin{table}[tbhp]
		\centering
		%\captionsetup{labelsep=period,labelfont=bf}
		\caption{Neural network training consistent hyperparameters.}
		\label{Table1}
		\begin{tabular}{*{6}{c}}
			\toprule
			Adam-iters&\qquad  L-BFGS-iters  &\qquad Adam-lr/L-BFGS-lr    &\qquad  $k$  \\
			\midrule
			
			$2000$  &\qquad $5000$ &\qquad  $0.001$  &\qquad $2$  \\
			% $40962$&    & &   & & &  \\
			%3.7841e-7}  & &   &  &  85.70\\
		\bottomrule
	    \end{tabular}
     \end{table}
\end{comment}

We compute the CEV model with $\beta=1$ in the experiment. In addition, the strike price, the asset volatility and the interest rate are set to be  $K=80$, $\sigma=0.3$ and $r=0.01$, respectively. The boundary condition used in the testing is 
$$U(0,t)=0, \ \  U(100,t)=Ke^{-rt}.$$

\begin{figure}[!htp]  
	\centering
	% 创建第一行第一列的子图  
	\begin{subfigure}{.19\textwidth}  
		\centering  
		\includegraphics[width=1\linewidth]{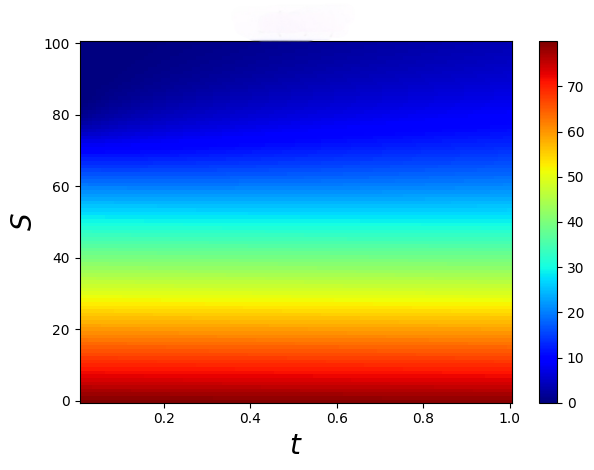}  
		\caption*{Exact}  
		\label{fig:16-1}  
	\end{subfigure}
	% 创建第一行第二列的子图  
	\begin{subfigure}{.19\textwidth}  
		\centering  
		\includegraphics[width=1\linewidth]{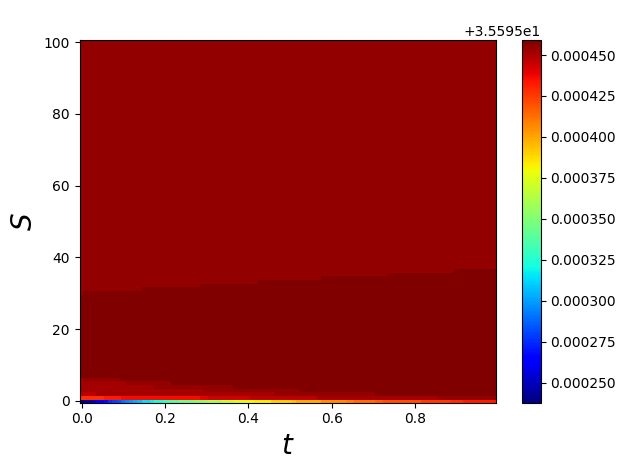} 
		\caption*{PINN}  
		\label{fig:16-2}  
	\end{subfigure}
        % 创建第一行第三列的子图 
	\begin{subfigure}{.19\textwidth}  
		\centering  
		\includegraphics[width=1\linewidth]{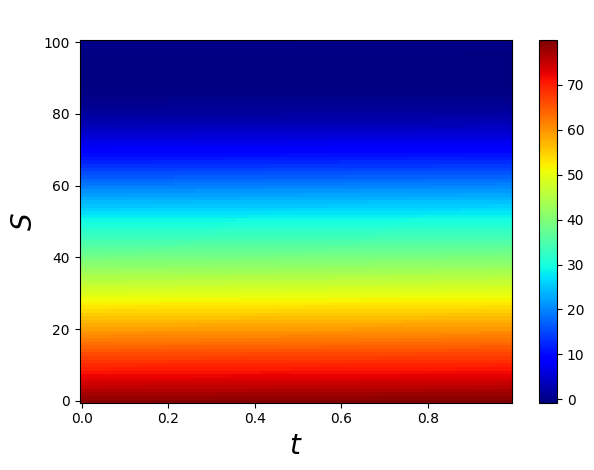}  
		\caption*{RAM-PINN}  
		\label{fig:16-3}  
	\end{subfigure}  
    % 创建第一行第三列的子图 
        \begin{subfigure}{.19\textwidth}  
		\centering  
		\includegraphics[width=1\linewidth]{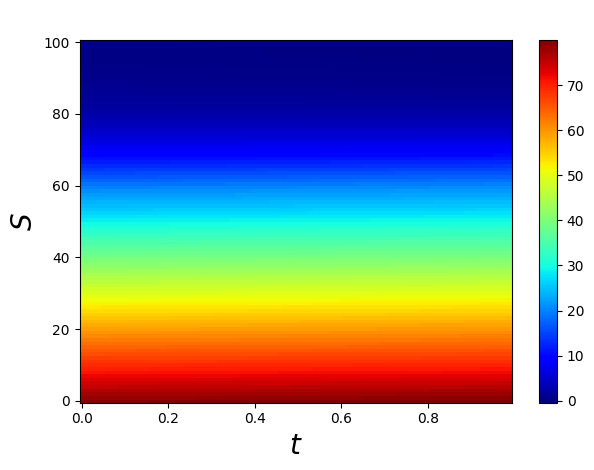}  
		\caption*{WAM-PINN}  
		\label{fig:16-5}  
	\end{subfigure}  
	% 创建第一行第三列的子图 
	\begin{subfigure}{.19\textwidth}  
		\centering  
		\includegraphics[width=1\linewidth]{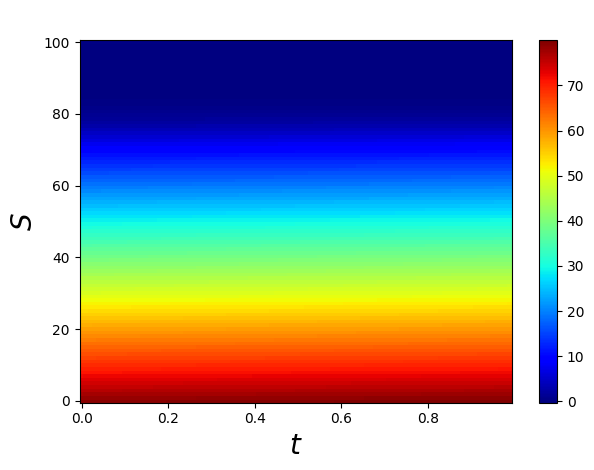}  
		\caption*{AM-PIRN}  
		\label{fig:16-4}  
	\end{subfigure}  

	% 为整个图像阵列添加标题和标签  
	\caption{The exact solution and the numerical solutions of  PINN, RAM-PINN, WAM-PINN and AM-PIRN for CEV equation.} % 整个图像阵列的标题  
	\label{fig:u_CEV} % 整个图像阵列的标签  
	
\end{figure}

In Fig. \ref{fig:u_CEV}, we show the numerical solutions of CEV model solved by PINN, RAM-PINN,  WAM-PINN and AM-PIRN with the same number of collocation points as Example 1, i.e., $|\mathcal{X}_i| = 100$, $|\mathcal{X}_b| = 200$, $|\mathcal{X}_{p,i}| = 1500$ and $|\mathcal{X}_{p,m}| = 500$, respectively. Compared to the former two models, the CEV equation is highly nonlinear duo to the local nonlinear volitility. So the PINN  provides the numerical solution  totally far away from the exact solution with the PDE residules reaching the order of $10^3$, while all the three adaptive sampling methods still give proper approximations. 
% The number of iteration rounds for RAM-PINN,AM-PIRN and WAM-PINN are both set to 10. For testing, a 100×100 uniform grid is used to compute the relative L2 error.}

As shown in Fig. \ref{fig18}, the PDE residuals of both RAM-PINN and WAM-PINN keep about the order of $10^{-1}$, while that of AM-PIRN yields a lower order of $10^{-2}$.  The largest residuals of RAM-PINN are located at the two ends of the time domain, but WAM-PINN and AM-PIRN show a band area of the larger residuals near the strike price $K=80$ which are more reasonable.  The absolute errors of the function estimation of all the adaptive sampling methods are quite close with the order of $10^{-2}$, but that of PINN is only $10^{-1}$. The large absolute errors of PINN occurred in the two boundaries, and there are higher absolute errors near the strike price $K=80$  for the adaptive methods. From the last row of the figure, we can see that the samples of AM-PIRN concentrated in the band area where both the large PDE 
residuals and large function absolute errors occurred.   

\begin{figure} [!htp]  
	\centering  
		\includegraphics[width=0.6\linewidth]{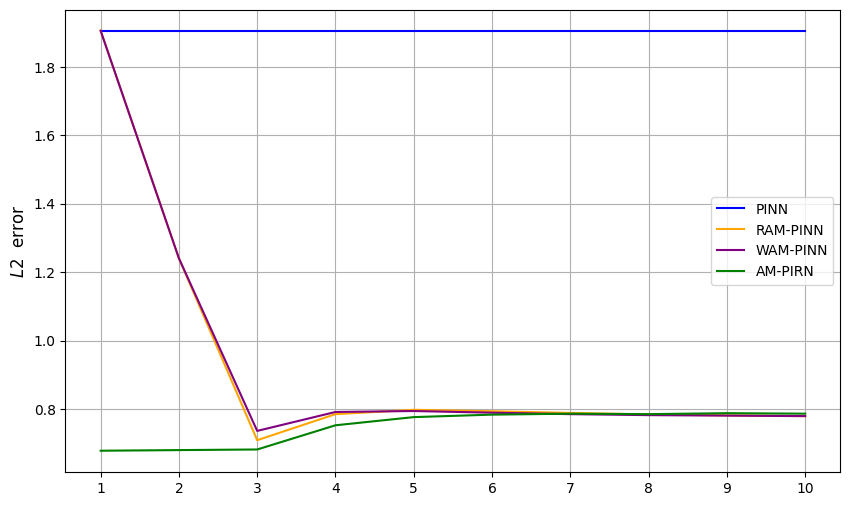}  
		%\label{fig:14-1}   
    	% 为整个图像阵列添加标题和标签  
	\caption{
    The iterative convergence curves of the $L_2$ errors (in log scale) of the estimated solutions for the CEV equation by PINN, RAM-PINN, WAM-PINN and AM-PIRN.} % 整个图像阵列的标题  
	\label{fig:CEV_convergence} % 整个图像阵列的标签
\end{figure}

\begin{figure}[!htp]  
	\centering  
	
	% 第一行
	\begin{subfigure}{.24\textwidth}  
		\centering  
		\includegraphics[width=.99\linewidth]{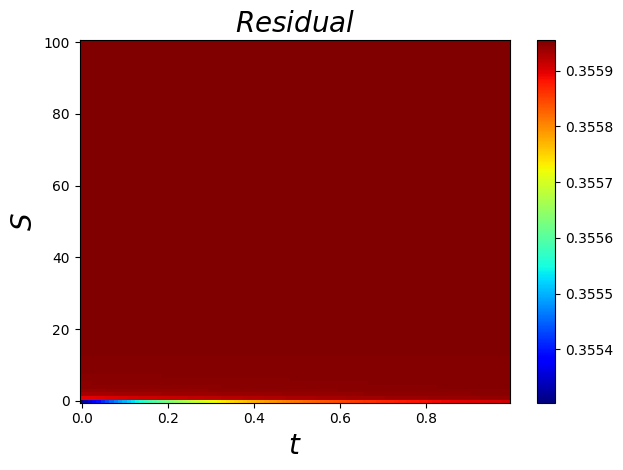} 
		%\caption*{PINN}  
		%\label{fig:18-1}  
	\end{subfigure}%  
	\begin{subfigure}{.24\textwidth}  
		\centering  
		\includegraphics[width=.99\linewidth]{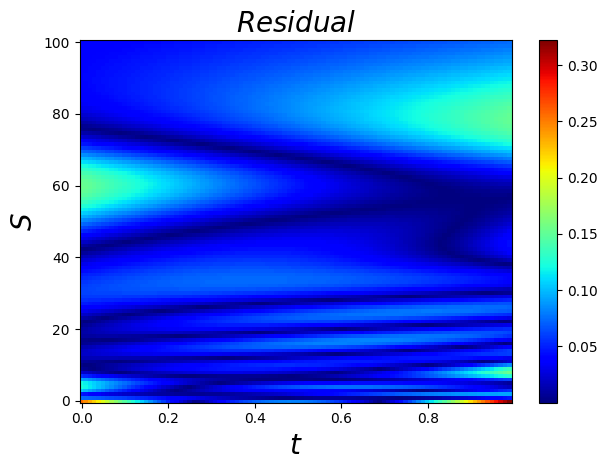} 
		%\caption*{AM-PINN}  
		%\label{fig:18-2}  
	\end{subfigure}%    
        \begin{subfigure}{.24\textwidth}  
		\centering  
		\includegraphics[width=.99\linewidth]{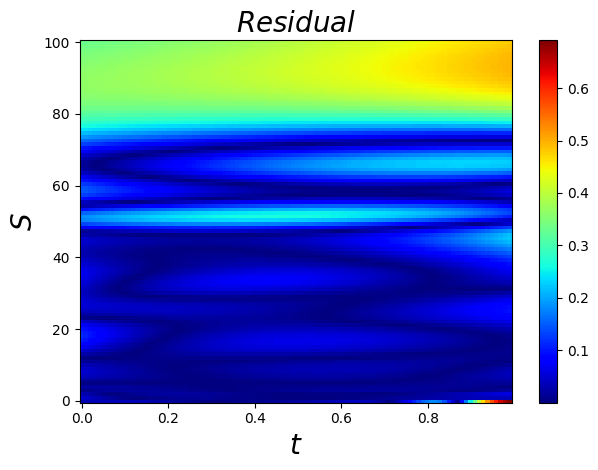} 
		%\caption*{AW-PIRN}  
		%\label{fig:18-3}  
	\end{subfigure}%
	\begin{subfigure}{.24\textwidth}  
		\centering  
		\includegraphics[width=.99\linewidth]{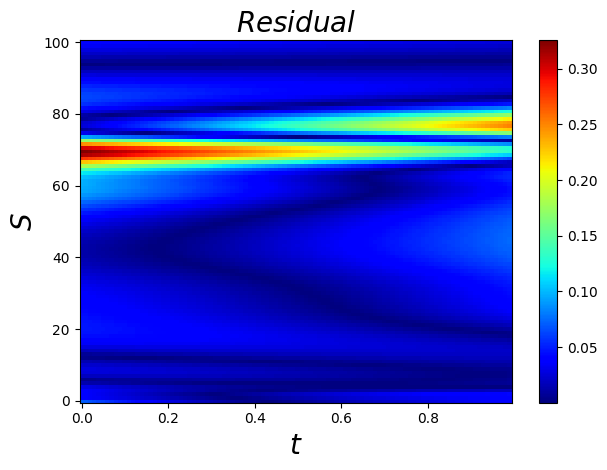} 
		%\caption*{AW-PIRN}  
		%\label{fig:18-3}  
	\end{subfigure}%

	% 第二行
	\begin{subfigure}{.24\textwidth}  
		\centering  
		\includegraphics[width=.99\linewidth]{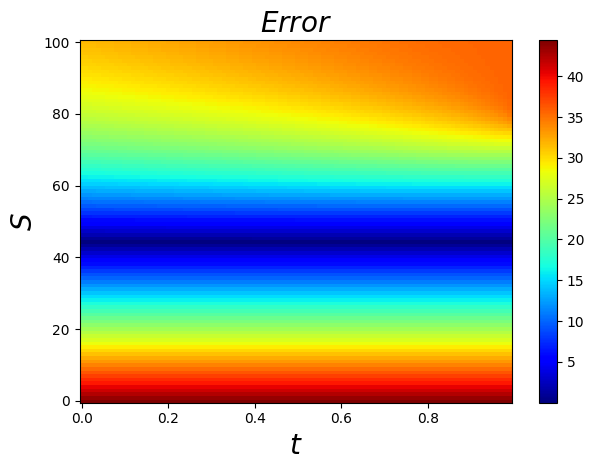}  
		%\caption*{PINN}  
		%\label{fig:18-5}  
	\end{subfigure}%  
	\begin{subfigure}{.24\textwidth}  
		\centering  
		\includegraphics[width=.99\linewidth]{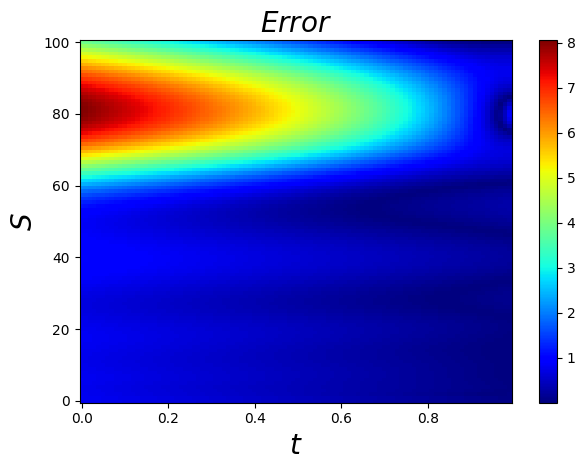}  
		%\caption*{AM-PINN}  
		%\label{fig:18-6}  
	\end{subfigure}% 
        \begin{subfigure}{.24\textwidth}  
		\centering  
		\includegraphics[width=.99\linewidth]{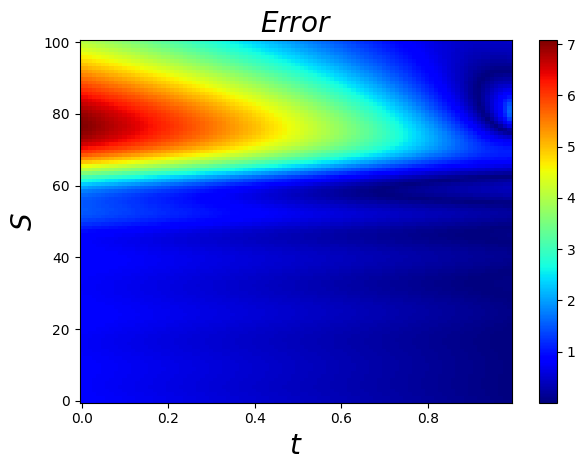} 
		%\caption*{AW-PIRN}  
		%\label{fig:18-8}  
	\end{subfigure}%
	\begin{subfigure}{.24\textwidth}  
		\centering  
		\includegraphics[width=.99\linewidth]{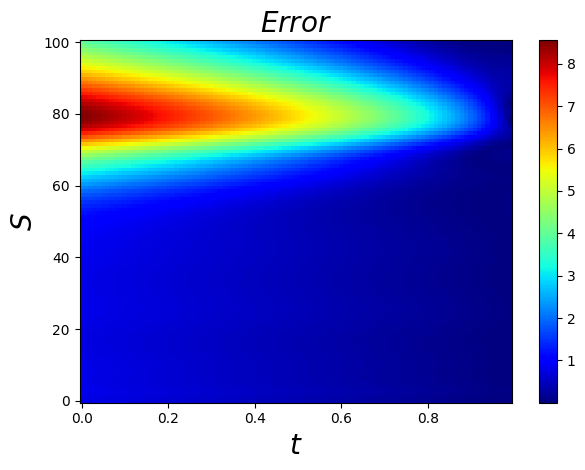}  
		%\caption*{AW-PIRN}  
		%\label{fig:18-7}  
	\end{subfigure}%

	% 第三行
	\begin{subfigure}{.24\textwidth}  
		\centering  
		\includegraphics[width=.99\linewidth]{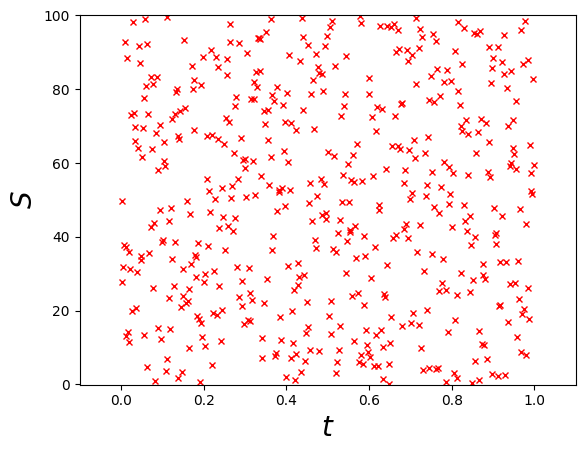} 
		\caption*{PINN}  
		\label{fig:18-9}  
	\end{subfigure}%  
	\begin{subfigure}{.24\textwidth}  
		\centering  
		\includegraphics[width=.99\linewidth]{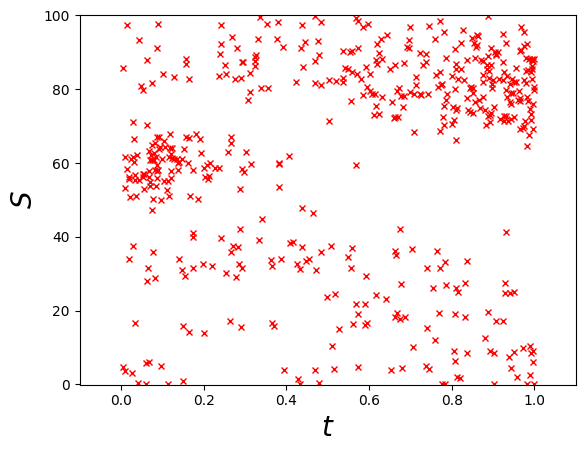}  
		\caption*{RAM-PINN}  
		\label{fig:18-10}  
	\end{subfigure}%  
        \begin{subfigure}{.24\textwidth}  
		\centering  
		\includegraphics[width=.99\linewidth]{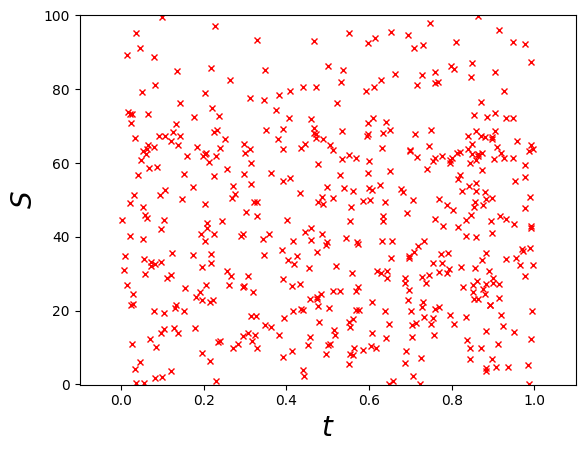} 
		\caption*{WAM-PINN}  
		\label{fig:18-12}  
	\end{subfigure}%
	\begin{subfigure}{.24\textwidth}  
		\centering  
		\includegraphics[width=.99\linewidth]{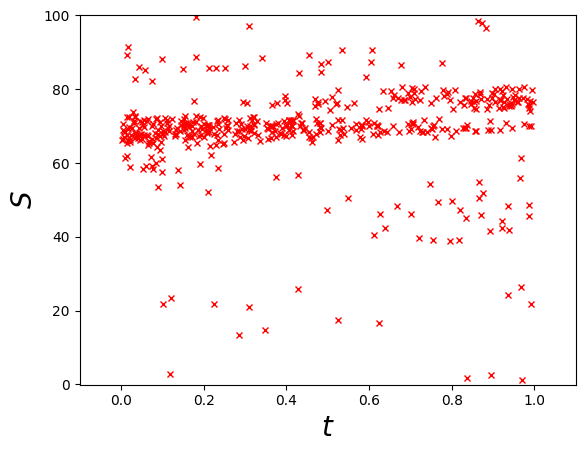}  
		\caption*{AM-PIRN}  
		\label{fig:18-11}  
	\end{subfigure}%
	% 为整个图像阵列添加标题和标签  
	\caption{
The PDE residuals (first row), absolute errors of estimated solutions (middle row), and distribution of mobile collocation points (last row) for  CEV model  computed by PINN, RAM-PINN, WAM-PINN and AM-PIRN after $10$ iteration rounds, respectively.}  
	\label{fig18} % 整个图像阵列的标签
\end{figure}

The convergence curves of the relative $L_2$ errors of all the methods with number of iteration rounds being $10$ are shown in Fig. \ref{fig:CEV_convergence}. It reveals that with the number of rounds for adaptive collocation points increasing,  the relative $L_2$ errors for RAM-PINN, WAM-PINN and AM-PIRN decrease to the order of $10^{-2}$ in the end. All the adaptive methods begin to convergent from the $6$-th round. However, compared to the curves of RAM-PINN and WAM-PINN, AM-PIRN decay faster at the beginning of the training.

At last in Table (\ref{Tab:CEV_1}) and (\ref{Tab:CEV_2}), we  test more experiments with different movable and immovable collocation points. If we keep the number of the movable points to be $500$ and increase the number of immovable points from $1000$ to $2000$, the PDE residuals of PINN reduced from $10^3$ to $10^{-1}$, but it still helps little to the option price estimation. AM-PIRN can show the accurate estimation even with $1000$ immovable points.  If we keep the number of stationary points being  $1500$, and increase the  the number of immovable points from $250$ to $750$, we can see that only AM-PIRN can achieve both accurate approximation of PDE and the option price with the order of $10^{-2}$.

% 导入图像阵列  

% 导入图像阵列  

\begin{table}[!htp]
    \centering
\begin{minipage}[c]{0.49\textwidth}
\centering
\footnotesize
\begin{tabular}{@{}lccc@{}}
\toprule
$|\mathcal{X}_{p,i}|$ & 1000 & 1500 & 2000 \\ \midrule
PINN & 2.29E$+$03 & 2.31E$+$03 & 3.60E$+$01 \\
RAM-PINN & 2.33E$-$01 & 2.82E$-$01 & 1.02E$-$01 \\
WAM-PINN & 1.69E$-$01 & 2.15E$-$01 & 4.76E$-$02 \\ 
AM-PIRN & 4.68E$-$02 & 4.98E$-$02 & 3.50E$-$02 \\ 
\bottomrule
\end{tabular}
\end{minipage}
\begin{minipage}[c]{0.49\textwidth}
\centering
\footnotesize
\begin{tabular}{@{}lccc@{}}
\toprule
$|\mathcal{X}_{p,i}|$ & 1000 & 1500 & 2000 \\ \midrule
PINN & 6.03E$-$01 & 6.06E$-$01 & 1.38E$-$01 \\
RAM-PINN & 5.35E$-$02 & 5.13E$-$02  & 5.52E$-$02 \\
WAM-PINN & 3.94E$-$02 & 5.29E$-$02  & 5.24E$-$02 \\ 
AM-PIRN & 4.11E$-$02 & 4.05E$-$02  & 5.64E$-$02 \\ 
\bottomrule
\end{tabular}
\end{minipage}
\caption{The PDE residuals (left) and the relative $L_2$ errors (right) of the CEV equation with $|\mathcal{X}_{p,m}| = 500$ movable points, respectively. } \label{Tab:CEV_1}
\end{table}

\begin{table}[!htp]
    \centering
\begin{minipage}[c]{0.49\textwidth}
\centering
\footnotesize
\begin{tabular}{@{}lccc@{}}
\toprule
$|\mathcal{X}_{p,m}|$ & 250 & 500 & 750 \\ \midrule
PINN & 2.43E$+$03 & 2.31E$+$03 & 6.62E$+$03 \\
RAM-PINN & 2.31E$+$03 & 2.82E$-$01 & 1.91E$-$01 \\
WAM-PINN & 2.31E$+$03 & 2.15E$-$01 & 2.31E$+$03 \\ 
AM-PIRN & 3.13E$-$02 & 4.98E$-$02 & 5.17E$-$02 \\ 
\bottomrule
\end{tabular}
%\caption{The residuals of CEV equation with $|\mathcal{X}_{p,i}| = 1500$ stationary points.}
\end{minipage}
\begin{minipage}[c]{0.49\textwidth}
\centering
\footnotesize
\begin{tabular}{@{}lccc@{}}
\toprule
$|\mathcal{X}_{p,m}|$ & 250 & 500 & 750 \\ \midrule
PINN & 6.14E$-$01 & 6.06E$-$01 & 1.21E$+$00 \\
RAM-PINN & 6.06E$-$01 & 5.13E$-$02 & 5.60E$-$02 \\
WAM-PINN & 6.06E$-$01 & 5.29E$-$02 & 6.06E$-$01 \\ 
AM-PIRN & 5.65E$-$02 & 4.05E$-$02 & 4.11E$-$02 \\ 
\bottomrule
\end{tabular}
\end{minipage}
\caption{The PDE residuals (left) and the relative $L_2$ errors (right) of CEV equation with $|\mathcal{X}_{p, i}| = 1500$ stationary points, respectively. } \label{Tab:CEV_2}
\end{table}

\vspace{3em}

%\\ \hspace*{\fill} \\

\noindent \textbf{Example 4: Heston equation}

In the last example, we will present the numerical results of  the  Heston stochastic volatility model,  which is one of the most popular two factor stochastic model. In the Heston model, the underlying asset $S_t$ follows a “geometric Brownian motion” with a stochastic volatility, and the square of the volatility follows a CIR process, i.e.,
\begin{align*}\label{Heston_S}
d S_t&=\mu S_t d t+\sqrt{v_t} S_t d W_t^1, \\
d v_t&=\kappa\left(\theta-v_t\right) d t+\sigma \sqrt{v_t} d W_t^2,
\end{align*}
where  $W_t^1$ and $W_t^2$ are correlated Brownian motions with the correlation given by $\rho$, $\mu$ is drift of the stock process,  $\kappa$ is reversion coefficient of the variance process, $\theta$ is long term mean of the variance process, and $\sigma$ is volatility coefficient of the variance process.

%\begin{align*}
%\frac{\partial V}{\partial t} + \frac{1}{2} f^2(Y) S^2 \frac{\partial^2 V}{\partial S^2} + \rho f(Y) \hat{\sigma} S \frac{\partial^2 V}{\partial S \partial Y} + \frac{1}{2} \hat{\sigma}^2 \frac{\partial^2 V}{\partial Y^2} + rS \frac{\partial V}{\partial S} + (\mu - \lambda \hat{\sigma}) \frac{\partial V}{\partial Y} - rV = 0
%\end{align*}

\begin{figure}[htbp]  
	\centering
	
	% 创建第一行第一列的子图  
	\begin{subfigure}{.19\textwidth}  
		\centering  
		\includegraphics[width=1\linewidth]{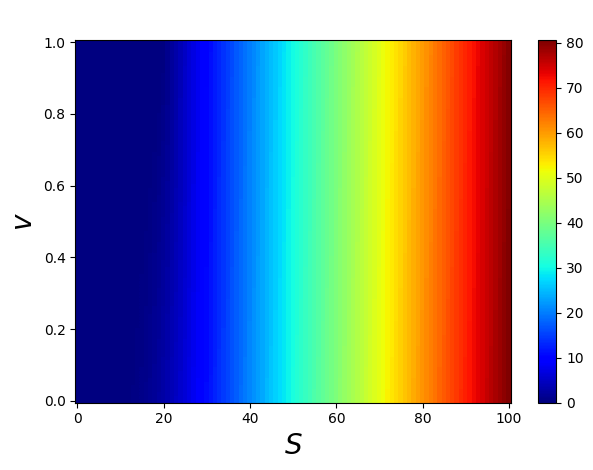}  
		\caption*{Reference}  
		\label{fig:19-1}  
	\end{subfigure}
	% 创建第一行第二列的子图  
	\begin{subfigure}{.19\textwidth}  
		\centering  
		\includegraphics[width=1\linewidth]{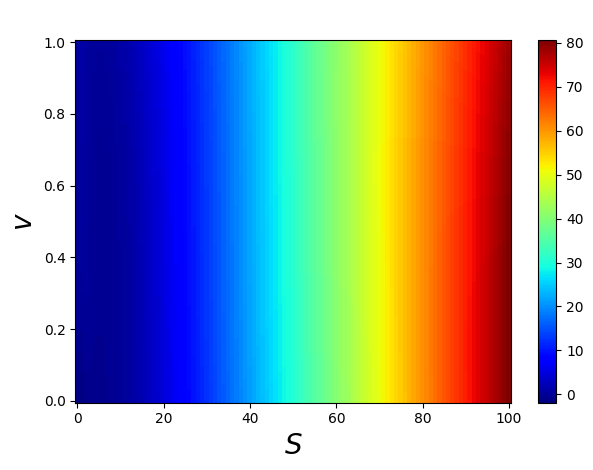} 
		\caption*{PINN}  
		\label{fig:19-2}  
	\end{subfigure}
        % 创建第一行第一列的子图  
	\begin{subfigure}{.19\textwidth}  
		\centering  
		\includegraphics[width=1\linewidth]{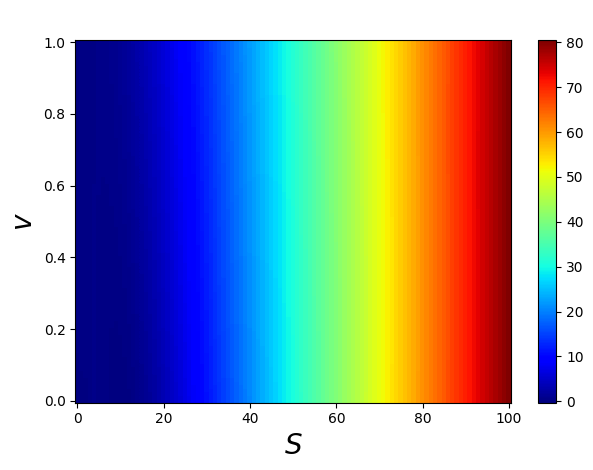}  
		\caption*{RAM-PINN}  
		\label{fig:19-3}  
	\end{subfigure}
        % 创建第一行第三列的子图 
	\begin{subfigure}{.19\textwidth}  
		\centering  
		\includegraphics[width=1\linewidth]{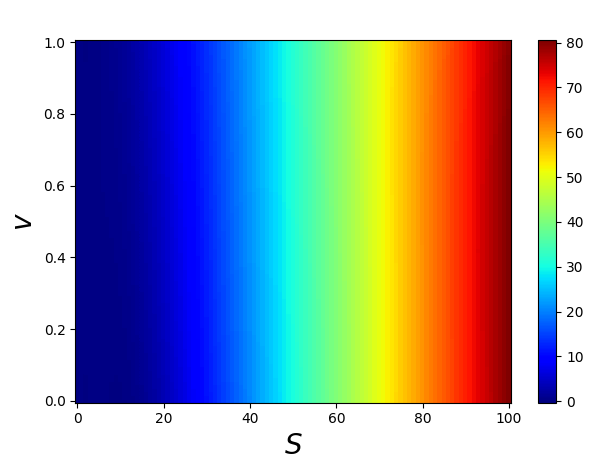}  
		\caption*{WAM-PINN}  
		%\label{fig:19-4}  
	\end{subfigure}  
        % 创建第一行第三列的子图 
	\begin{subfigure}{.19\textwidth}  
		\centering  
		\includegraphics[width=1\linewidth]{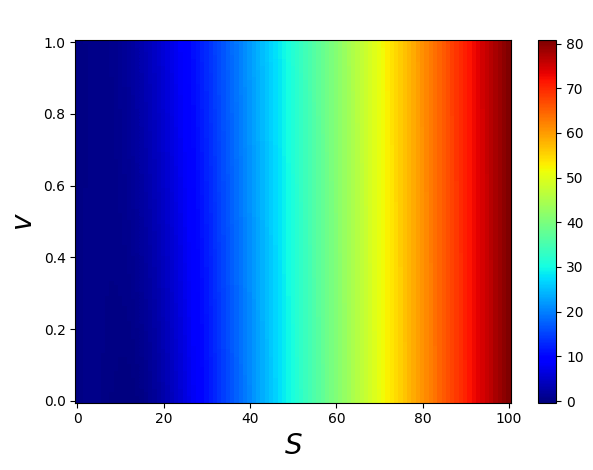}  
		\caption*{AM-PIRN}  
		%\label{fig:19-4}  
	\end{subfigure}  

	% 为整个图像阵列添加标题和标签  
	\caption{The reference solution and the numerical solutions of Heston equation at $t=0$ computed by PINN, RAM-PINN, WAM-PINN and AM-PIRN.} % 整个图像阵列的标题  
	\label{fig:u_heston} % 整个图像阵列的标签  
	
\end{figure}

Following the arbitrage arguments similar to the one mentioned in Section 2, we can obtain the following equation for the option price $U(S,v,t)$ of European call option, i.e., 
{\small 
\begin{align*}
\begin{cases}
\frac{\partial U}{\partial t} + \frac{1}{2} v S^2 \frac{\partial^2 U}{\partial S^2} + \rho \sigma v S \frac{\partial^2 U}{\partial S \partial v} + \frac{1}{2} \sigma^2 v \frac{\partial^2 U}{\partial v^2} + r S \frac{\partial V}{\partial S} + [\kappa(\theta - v) - \lambda(S,v,t)] \frac{\partial U}{\partial v} - r U = 0,   \\
U(S, v, T) = (S - K)^+, 
\end{cases}
\end{align*}
}
with $\lambda(S,v,t)$ being the price of volatility risk. 
For simplicity, we will set $\lambda(S,v,t)=0$ in the experiment.  A closed-form solution for this model using characteristic function has been derived in \cite{Heston}, and the reference solution provided in this example uses numerical computation to obtain the integral in the closed-form solution. The  test parameters are given by $K=20$, $r=0.03$, $\rho=0.8$, $\theta=0.2$, $\sigma=0.3$ and $\kappa=2$. 
Compared to the above examples, the Heston is a multi-dimensional equation. So  
the training dataset is increased to 1000 initial points, 2000 boundary points, and 20000 collocation points uniformly distributed in the computational domain, which includes 15000 stationary collocation points  and 5000 adaptive collocation points. 
We show the reference solutions of Heston equation  with numerical solutions of PINN, RAM-PINN, WAM-PINN and AM-PIRN in Fig. \ref{fig:u_heston} at $t=0$. 

\begin{figure}[!htp]  
	\centering  
	
	% 第一行
        \begin{subfigure}{.24\textwidth}  
		\centering  
		\includegraphics[width=.99\linewidth]{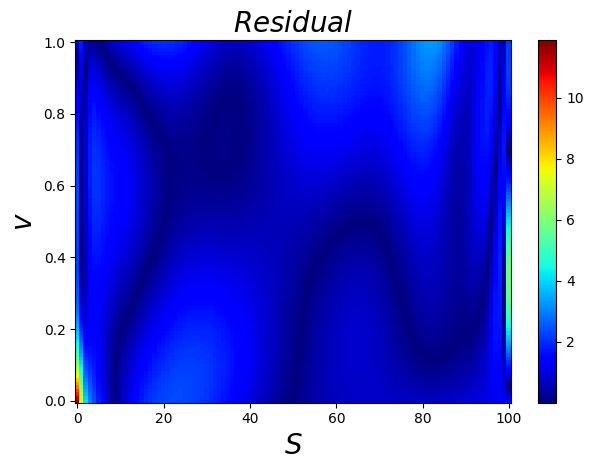}  
		%\caption*{PINN}  
		\label{fig:20-1}  
	\end{subfigure}%  
        \begin{subfigure}{.24\textwidth}  
		\centering  
		\includegraphics[width=.99\linewidth]{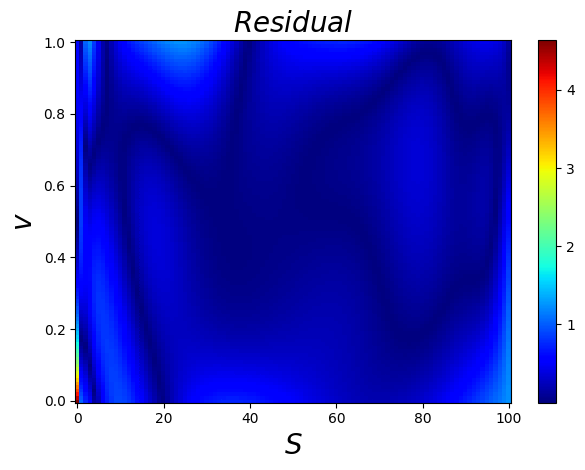}  
		%\caption*{AM-PINN}  
		\label{fig:20-2}  
	\end{subfigure}%  
        \begin{subfigure}{.24\textwidth}  
		\centering  
		\includegraphics[width=.99\linewidth]{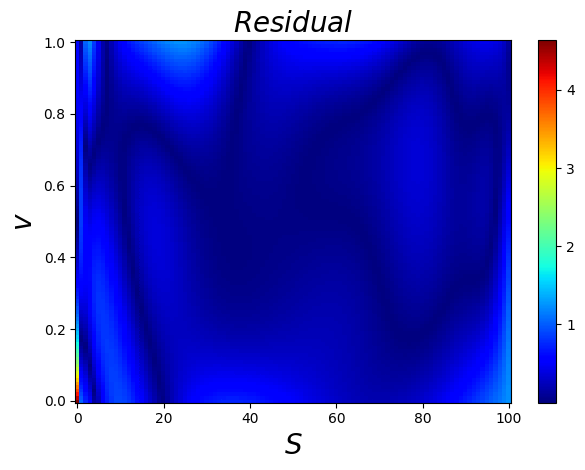}  
		%\caption*{AW-PIRN}  
		\label{fig:20-3}  
	\end{subfigure}%
	\begin{subfigure}{.24\textwidth}  
		\centering  
		\includegraphics[width=.99\linewidth]{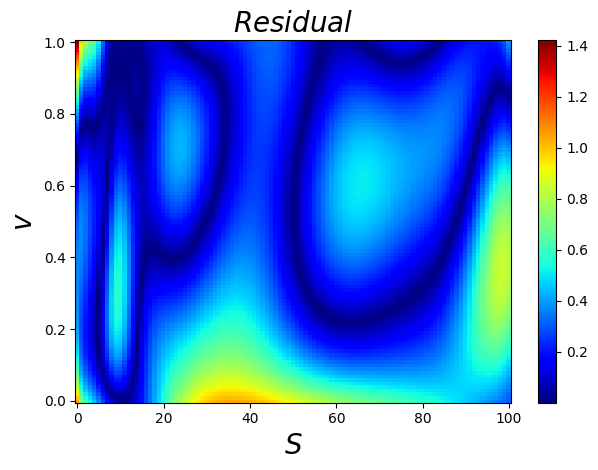}  
		%\caption*{AW-PIRN}  
		%\label{fig:20-3}  
	\end{subfigure}%

	\begin{subfigure}{.24\textwidth}  
		\centering  
		\includegraphics[width=.99\linewidth]{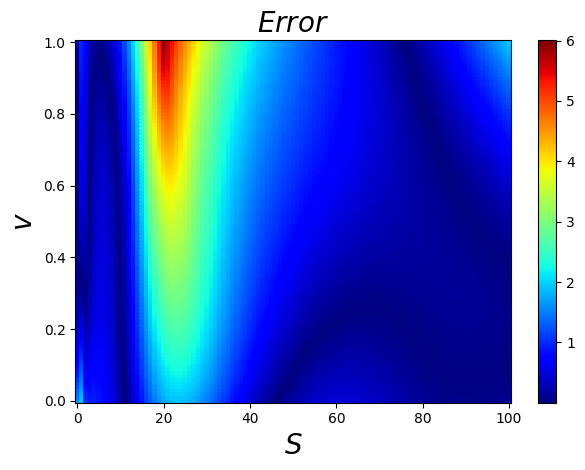}  
		%\caption*{PINN}  
		\label{fig:20-4}  
	\end{subfigure}%  
        \begin{subfigure}{.24\textwidth}  
		\centering  
		\includegraphics[width=.99\linewidth]{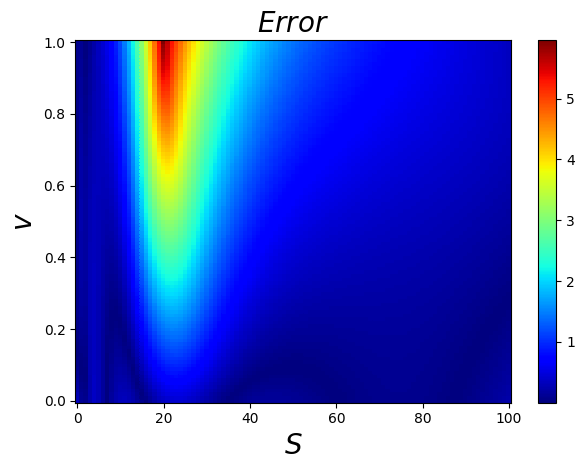}  
		%\caption*{AM-PINN}  
		\label{fig:20-5}  
	\end{subfigure}%  
        \begin{subfigure}{.24\textwidth}  
		\centering  
		\includegraphics[width=.99\linewidth]{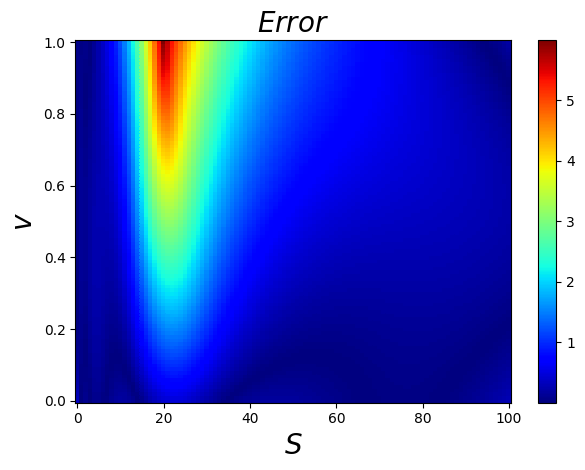}  
		%\caption*{AW-PIRN}  
		%\label{fig:20-6}  
	\end{subfigure}%
	\begin{subfigure}{.24\textwidth}  
		\centering  
		\includegraphics[width=.99\linewidth]{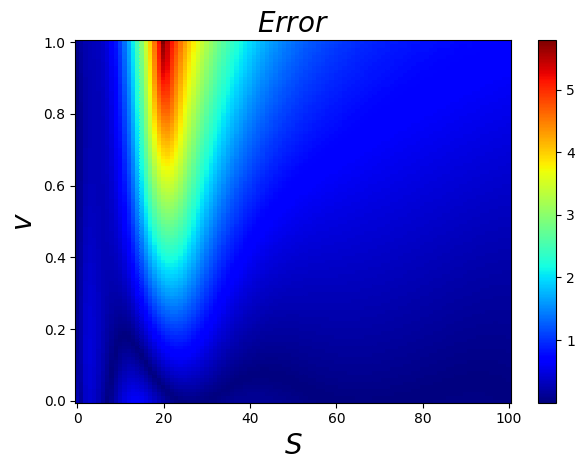}  
		%\caption*{AW-PIRN}  
		\label{fig:20-6}  
	\end{subfigure}%

		\begin{subfigure}{.24\textwidth}  
		\centering  
		\includegraphics[width=.99\linewidth]{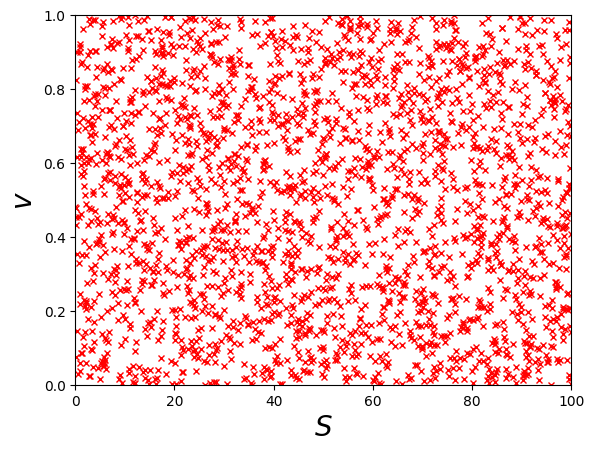}  
		\caption*{PINN}  
		\label{fig:20-7}  
	\end{subfigure}%  
        \begin{subfigure}{.24\textwidth}  
		\centering  
		\includegraphics[width=.99\linewidth]{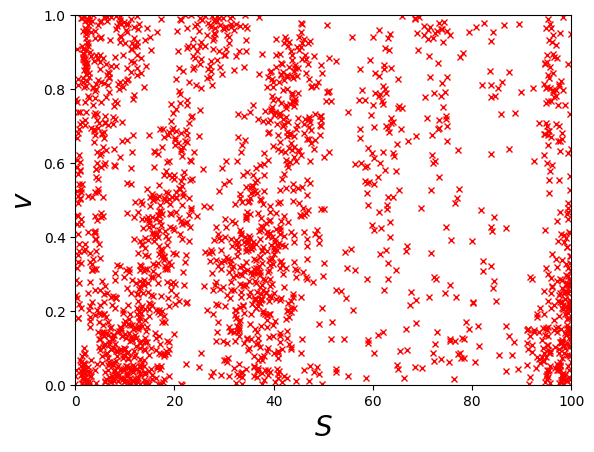}  
		\caption*{RAM-PINN}  
		\label{fig:20-8}  
	\end{subfigure}%  
        \begin{subfigure}{.24\textwidth}  
		\centering  
		\includegraphics[width=.99\linewidth]{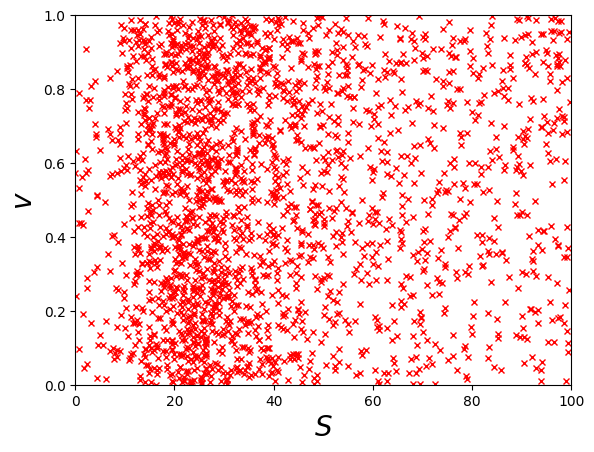}  
		\caption*{WAM-PINN}  
		%\label{fig:20-9}  
	\end{subfigure}%
	\begin{subfigure}{.24\textwidth}  
		\centering  
		\includegraphics[width=.99\linewidth]{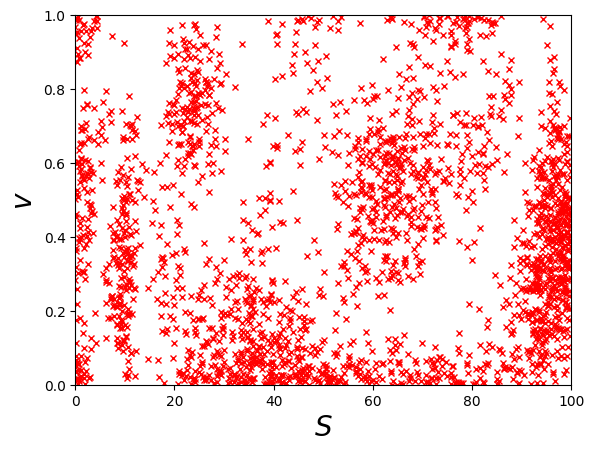}  
		\caption*{AM-PIRN}  
		\label{fig:20-9}  
	\end{subfigure}%

	% 为整个图像阵列添加标题和标签  
	\caption{
    The PDE residuals (first row), absolute errors of estimated solutions (middle row), and distribution of mobile collocation points (last row) for the Heston equation   computed by PINN, RAM-PINN, WAM-PINN and AM-PIRN after $10$ iteration rounds, respectively. (The time domain of the sampling points is from  $t=0$ to $t=0.5$.)}  
	\label{fig: Heston_res} % 整个图像阵列的标签
\end{figure}

As shown in Fig. \ref{fig: Heston_res}, at $t=0$, the PDE residuals of PINN keep about the order of $10^{0}$, while those of RAM-PINN, WAM-PINN and AM-PIRN yield a lower order of $10^{-1}$.  The largest residuals of all the methods are located at the boundaries of the domain. The larger absolute errors of all the methods occurred near the strike price $K=20$. The absolute errors of the function estimation of all the adaptive sampling methods are quite close with the order of $10^{-2}$, but those of PINN is only $10^{-1}$.  In the last row of the figure, we show the sampling points in the time domain $[0,0.5]$. 
It can been seen that the more samples of WAM-PINN are located in a band area of near the strike price, since the resampling  density function of WAM-PINN uses the gradient of the option pricing function.  The samples of RAM-PINN using the  PDE residuals in the resampling density function  are dense in the larger residual area, but they do not show the concentration in the large function error domain.  
AM-PIRN combines the resampling density function with the PDE loss and the Resnet, which makes the samples concentrated in the area where both the large PDE residuals and large function absolute errors occurred. 

\begin{figure} [!htp]  
	\centering  
		\includegraphics[width=0.6\linewidth]{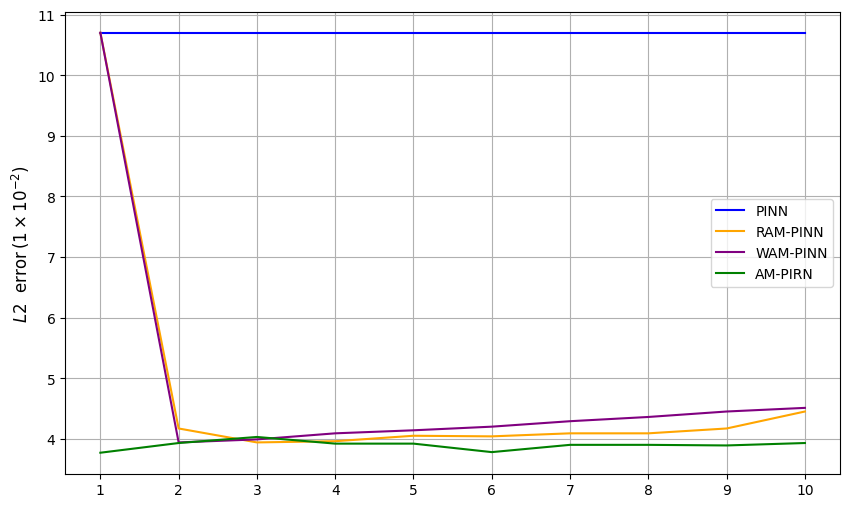}  
		%\label{fig:14-1}   
    	% 为整个图像阵列添加标题和标签  
	\caption{The iterative convergence curves of the $L_2$ errors of the estimated solutions for the Heston equation by PINN, RAM-PINN, WAM-PINN and AM-PIRN.
    } % 整个图像阵列的标题  
	\label{fig:HESTON_convergence} % 整个图像阵列的标签
\end{figure}

Fig. \ref{fig:HESTON_convergence} compares the convergence of relative $L_2$ errors (scaled by $10^2$) for PINN, RAM-PINN, WAM-PINN, and AM-PIRN in solving the Heston equation. AM-PIRN achieves stable convergence by the 8-th iteration round, demonstrating faster error reduction compared to other methods. In contrast, RAM-PINN and WAM-PINN exhibit a slight increase in $L_2$ error after the 10-th iteration round, highlighting challenges in maintaining stability during prolonged training.  

The numerical performance of collocation strategies for the Heston stochastic volatility model is systematically evaluated in  Tables (\ref{Tab:Heston_1}) and (\ref{Tab:Heston_2}). When maintaining $|\mathcal{X}_{p,m}| = 5000$ movable points while increasing immovable collocation points from $10000$ to $20000$, PINN demonstrates PDE residuals decreasing from $4.00 \times 10^{0}$ to $2.29 \times 10^{0}$, with corresponding relative $L_2$ errors ranging $5.76 \times 10^{-2}$ to $6.54 \times 10^{-2}$. Comparatively, adaptive methods exhibit smaller PDE residuals (about $10^{-1}$) and achieve superior option price error stability, maintaining $L_2$ errors below $3.93 \times 10^{-2}$ across all configurations.

For the reverse configuration with fixed $|\mathcal{X}_{p,i}| = 15000$ and variable movable points (from $2500$ to $7500$), PINN shows residual reduction from $7.02 \times 10^{0}$ to $4.95 \times 10^{0}$ as $|\mathcal{X}_{p,m}|$ increases, while WAM-PINN and AM-PIRN achieves optimal performance at $|\mathcal{X}_{p,m}| = 7500$ with residual $5.39 \times 10^{-1}$ (WAM-PINN) and $L_2$ error $3.86 \times 10^{-2}$ (AM-PIRN). The experimental data confirms that adaptive collocation methods provide error control outperforming PINN's minimum observed error.

%Then in Table (\ref{Tab:Heston_1}) and (\ref{Tab:Heston_2}), we test the Heston equation with more experiments by changing the movable and immovable collocation points. If we keep the number of the movable points to be $5000$ and increase the number of immovable points from $10000$ to $20000$, the PDE residual  and the $L_2$ error of option pricing function  of PINN are both one order lower than the adaptive sampling methods.  If we keep the number of stationary points being  $15000$, and increase the number of immovable points from $2500$ to $7500$, we can see the similar results.

% 导入图像阵列  

% 导入图像阵列  

\begin{table}[!htp]
    \centering
\begin{minipage}[c]{0.49\textwidth}
\centering
\footnotesize
\begin{tabular}{@{}lccc@{}}
\toprule
$|\mathcal{X}_{p,i}|$ & 10000 & 15000 & 20000 \\ \midrule
PINN & 4.00E$-$00 & 5.46E$-$00 & 2.29E$-$00 \\
RAM-PINN & 5.20E$-$01 & 4.89E$-$01 & 5.93E$-$01 \\
WAM-PINN & 5.03E$-$01 & 4.29E$-$01 & 4.02E$-$01 \\ 
AM-PIRN & 6.14E$-$01 & 5.91E$-$01 & 5.67E$-$01 \\ 
\bottomrule
\end{tabular}
%\caption{The residuals for Heston equation with $|\mathcal{X}_{p,m}| = 5000$ movable points.}
\end{minipage}
\begin{minipage}[c]{0.49\textwidth}
\centering
\footnotesize
\begin{tabular}{@{}lccc@{}}
\toprule
$|\mathcal{X}_{p,i}|$ & 10000 & 15000 & 20000 \\ \midrule
PINN & 5.76E$-$02 & 1.07E$-$01 & 6.54E$-$02 \\
RAM-PINN & 3.96E$-$02& 3.94E$-$02 & 3.85E$-$02  \\
WAM-PINN & 3.98E$-$02 & 3.95E$-$02 & 3.91E$-$02 \\
AM-PIRN & 3.93E$-$02 & 3.78E$-$02 & 3.84E$-$02 \\ 
\bottomrule
\end{tabular}
\end{minipage}
\caption{The PDE residuals (left) and the relative $L_2$ errors (right) of Heston equation with $|\mathcal{X}_{p,m}| = 5000$ movable points, respectively.}\label{Tab:Heston_1}
\end{table}

\begin{table}[!htp]
    \centering
\begin{minipage}[c]{0.49\textwidth}
\centering
\footnotesize
\begin{tabular}{@{}lccc@{}}
\toprule
$|\mathcal{X}_{p,m}|$ & 2500 & 5000 & 7500 \\ \midrule
PINN & 7.02E$-$00 & 5.46E$-$00 & 4.95E$-$00 \\
RAM-PINN & 4.69E$-$01 & 4.89E$-$01 & 5.46E$-$01 \\
WAM-PINN & 4.83E$-$01 & 4.29E$-$01 & 5.39E$-$01 \\ 
AM-PIRN & 6.01E$-$01 & 5.91E$-$01 & 6.74E$-$01 \\
\bottomrule
\end{tabular}
%\caption{The residuals for Heston equation with $|\mathcal{X}_{p,i}| = 15000$ stationary points.}
\end{minipage}
\begin{minipage}[c]{0.49\textwidth}
\centering
\footnotesize
\begin{tabular}{@{}lccc@{}}
\toprule
$|\mathcal{X}_{p,m}|$ & 2500 & 5000 & 7500 \\ \midrule
PINN & 4.15E$-$02 & 1.07E$-$01 & 6.54E$-$02 \\
RAM-PINN & 4.02E$-$02 & 3.94E$-$02 & 3.91E$-$02 \\
WAM-PINN & 4.06E$-$02 & 3.95E$-$02 & 3.86E$-$02\\ 
AM-PIRN & 3.81E$-$02 & 3.78E$-$02 & 3.85E$-$02  \\ 
\bottomrule
\end{tabular}
%\caption{The relative $L_2$ error for Heston equation with $|\mathcal{X}_{p,i}| = 15000$ stationary points.}
\end{minipage}
\caption{
The PDE residuals (left) and the relative $L_2$ errors (right) of Heston equation with $|\mathcal{X}_{p,i}| = 15000$ stationary points, respectively. 
}\label{Tab:Heston_2}
\end{table}

\section{Conclusions}
In this paper, the proposed AM-PIRN framework effectively addresses the challenges of nonlinear option pricing PDEs by integrating adaptive samples redistribution based on PDE residuals with a ResNet architecture, achieving high accuracy in scenarios with sharp gradients and multi-dimensional complexity.  
In the last section, the numerical experiments on models such as the generalized Black-Scholes equation and multi-dimensional Heston equation demonstrate AM-PIRN’s superiority over PINN, RAM-PINN, and WAM-PINN, particularly in multi-dimensional or complex cases where adaptive sampling and architectural robustness excel.

%Numerical experiments of nonlinear option pricing models, such as general Black-sholes equation and multi-dimentional Heston equation have been used to show that AM-PIRN outperforms PINN, RAM-PINN, and WAM-PINN in both resolving PDE constraints and accurately estimating option prices. 
%The method’s advantages are particularly pronounced in complex or multi-dimensional models, where its adaptive sampling and robust architecture effectively mitigate challenges posed by sharp gradients and high nonlinearity.

Our future work will extend AM-PIRN to broader high-dimensional option pricing models, including stochastic volatility and path-dependent derivatives, while optimizing its adaptive sampling algorithms for real-time trading applications. Additional directions include enhancing computational efficiency for large-scale portfolios and integrating uncertainty quantification to strengthen reliability in dynamic market environments.

%This paper introduces a novel approach, combining Physically Informed ResNets (PIRNs) with adaptive sampling, to solve nonlinear Black-Scholes (BS) equations. The proposed RAR-PIRN method enhances traditional Physically Informed Neural Networks (PINNs) by replacing the fully connected neural network with a ResNet architecture and incorporating an adaptive sampling technique. These improvements not only increase the accuracy of solutions for complex BS equations but also stabilize the network by addressing the uneven distribution of training points.

%In experimental evaluations, the RAR-PIRN method achieves a significant reduction in relative error, decreasing by an order of magnitude compared to standard PINN and RAR-PINN methods. Furthermore, this study reveals that residuals do not always correspond to errors, emphasizing the need for future research to develop metrics that better capture the underlying physical properties of equations, moving beyond reliance on residual constraints in PINNs. Additionally, we believe this method has broader applicability to other scenarios, which will our future work.

\section*{Declarations}

\noindent {\bf Ethics approval and consent to participate} \ Not applicable.
\vspace{3mm}
\quad \\
{\bf Consent for publication} \  Not applicable.
\vspace{3mm}
\quad \\
{\bf Funding} This work is supported by the National Natural Science Foundation of China (Grants No. 12001487 and No. 11901377),  and the Characteristic  $\&$  Preponderant Discipline of Key Construction Universities in Zhejiang Province (Zhejiang Gongshang University- Statistics).
\vspace{3mm}
\quad \\
{\bf Data availability} \ The data will be available on request to the authors. 
\vspace{3mm}
\quad \\
\noindent {\bf Competing interests} \  The authors declare that they have no competing interests. 
\vspace{3mm}
\quad \\
\noindent {\bf Authors' contributions} \ 
Qinjiao Gao: Conceptualization, Methodology, Writing original draft. 
Zuowei Wang: Writing review $\&$ editing, Software.
Ran Zhang: Validation, Writing review $\&$ editing, Supervision.
Dongjiang Wang: Writing review $\&$ editing, Software.
All authors read and approved the final manuscript.
\vspace{3mm}
\quad \\
\noindent {\bf Acknowledgments} \ Not applicable.

\bibliographystyle{plain} % 按引用顺序排序
\bibliography{reference} % reference.bib 文件
\end{document}